\documentclass[preprint,12pt]{elsarticle}

\usepackage{amssymb}
\usepackage{amsmath}
\usepackage{xspace}
\usepackage{subfig}
\usepackage{color}
\usepackage{graphicx}
\usepackage{multirow}
\usepackage{lineno}
\usepackage{units}
\usepackage[hidelinks]{hyperref}
\usepackage{cleveref}

\crefname{figure}{Fig.}{Figs.}
\crefname{equation}{Eq.}{Eqs.}
\crefname{section}{Sec.}{Secs.}
\crefname{appendix}{Appendix}{Appendices}

\def \xmax {$X_{\rm{max}}$\xspace}

\def \rmu{$r_{\mu}$\xspace}
\def \rmumean{$\langle r_{\mu} \rangle$\xspace}
\def \deltarmu{$\Delta r_{\mu}$\xspace}
\def \deltarmumean{$\Delta \langle r_{\mu} \rangle$\xspace}
\def \dx{$DX$\xspace}

\def \smutop{$S_{\mu}^{\rm{sur}}$\xspace}
\def \smubur{$S_{\mu}^{\rm{bur}}$\xspace}

\def \sibyll{Sibyll2.3\xspace}
\def \qgsjet{QGSJetII-04\xspace}
\def \epos{EPOS-LHC\xspace}
\def \fluka{FLUKA\xspace}
\def \gheisha{GHEISHA\xspace}
\def \urqmd{UrQMD\xspace}

\journal{Astroparticle Physics}

\begin{document}


\begin{frontmatter}

\title{A new air-shower observable to constrain hadronic interaction models}

\author[1]{Raul R. Prado}
\author[1]{Vitor de Souza}
\address[1]{Instituto de F\'isica de S\~ao Carlos, Universidade de S\~ao Paulo, S\~ao Carlos, Brazil}

\begin{abstract}

The energy spectrum of muons at ground level in air showers is studied and a new observable is proposed to constrain hadronic interaction models used in air shower simulations. An asymmetric Gaussian function is proposed to describe the muon ground energy spectrum and its parameters are studied regarding primary particle, energy and hadronic interaction models. Based on two realistic measurements of the muon density at a given distance from the shower axis, a new observable (\rmu) is defined. Considering realistic values of detector resolutions and number of measured events, it is also shown \rmu can be successfully used to constrain low and high energy hadronic interaction models. The study is focused in the energy range between $10^{17.5}$ and $10^{18.0}$ eV because of the importance of this interval for particle physics and astrophysical models. The constraining power of the new observable is shown to be large within current experimental capabilities.

\end{abstract}


\end{frontmatter}

\section{Introduction}
\label{sec:intro}

The interaction of ultra high energy cosmic rays with atmospheric nuclei allows us to access hadronic interactions at energies much beyond the reach of man-made accelerators. However several properties of the cosmic ray phenomena and of the detection techniques impose limitations on the available information to study the highest energetic interactions. Experiments are only able to measure the air shower produced by the interaction of the cosmic particles with the nucleus of atoms in the atmosphere. The study of elementary properties of particle physics is done by relating global shower parameters to the properties of the hadronic interactions.

The depth at which the shower reaches its maximum number of particles (\xmax) and the muon component are the most used shower features related to properties of particle interactions. At the same time, they are also strongly sensitive to the type of the shower primary particle~\cite{Kampert2012}. Given that the primary particle type cannot be determined on event-by-event basis, different primary compositions must be considered in the study of the interaction properties. Usually, the large possibility of primary particles, from proton to iron nuclei, makes it impossible to disentangle the mass of the cosmic particle from particle interaction properties.

This paper proposes an analysis procedure to constrain hadronic interaction models by the measurements of the muon density by two different detectors. The idea proposed here is to compare the predictions of the models to measurements based on the muon density at ground level. It will be shown in~\cref{sec:obser} that the muon ground energy spectrum has valuable information, which allow to discriminate among different hadronic interaction models.

Hadronic interaction models used in Monte Carlo simulations of air showers are limited to phenomenological approaches tested and tuned to collider data ~\cite{Engel2011}. Three hadronic models for high energy interaction ($> 80$ GeV) and three hadronic models for low energy interaction ($< 80$ GeV) were investigated. A new and realistic air-shower observable based on two measurements of the muon density by different detectors is proposed in this paper and it is demonstrated to be powerful enough to constrain the hadronic interactions models.

The analysis procedure proposed here does not aim to infer properties of the particle interactions, instead, the new proposed parameter has power to test the predictions of the hadronic interaction models within realistic experimental conditions. By comparing the prediction of the models to future measurements of the new parameter, it will be possible to select the best model and guide the way towards a better understanding of the underlying assumptions taken by that model.

The work is based on Monte Carlo simulations of air showers. The new proposed parameter is studied in the energy range from $10^{17.5}$ to $10^{18.0}$ eV. The energy range is limited in order to minimize the effect of systematic uncertainties in energy reconstruction and also because of the importance of this range, which is the transition energy range from collider data to the extrapolation domain. It is shown that the new parameter can be used to test hadronic interaction models without knowledge on the primary composition.

The paper is organized as follows: \cref{sec:sim} describes the simulations, \cref{sec:carac} shows a study on the energy spectra of muons at the ground level, \cref{sec:obser} proposes the new parameter and studies its behavior with relation to primary mass, energy and detector properties, \cref{sec:results} quantifies the constraining power of the new parameter and~\cref{sec:conclusion} concludes the work.
\section{Simulations}
\label{sec:sim}

Extensive air showers were simulated using the Monte Carlo code CORSIKA v7.500~\cite{Heck1998a}. The applied thinning factor was $10^{-2}$ for the electromagnetic component and $10^{-4}$ for the hadronic component, with the maximum weight for any particle set to $100$. It was verified that the choice of this thinning configuration does not result in a bias in the present analysis. The energy of the primary particles was sampled continuously between $10^{17.5}$ and $10^{18.0}$ eV, following a power law energy spectrum with index -3. The arrival directions were sampled following a uniform distribution in solid angle, up to a maximum zenith angle of $60^{\circ}$. Three primaries were simulated: proton, nitrogen nucleus and iron nucleus. The minimum energy of particles simulated in air showers were set to $0.3$ GeV for hadrons and muons, and $0.003$ GeV for electrons and photons. Three hadronic interaction models were used for high energies interactions ($> 80$ GeV): \qgsjet~\cite{Ostapchenko2010}, \epos~\cite{Werner2007,Pierog2013} and \sibyll~\cite{Riehn2015}, and three for low energies ($< 80$ GeV): \fluka~\cite{Fluka}, \gheisha~\cite{GHEISHA1985} and \urqmd~\cite{Bleicher1999}. For each selected combination of hadronic interaction model and primaries particle, 1200 showers were generated. The ground altitude (1400 m above sea level) was chosen to be the one at the Pierre Auger Observatory.

The energy spectrum of muons at ground level is the shower feature to be considered in this paper. The energy spectra were built by collecting from the simulated air showers all muons reaching ground in a lateral distance between $425$ and $475$ meters from the shower axis. Although a full detector reconstruction is not applied, in~\cref{sec:obser:detector} the detector effects are taken into account by artificial smearing the shower observables around its simulated values. Detector thresholds and geometry are considered as well.

To sample showers at the same stage of development, the commonly used \dx parameter is used. \dx is the slant atmospheric depth between the shower maximum and the ground. It is given by

\begin{equation}
  DX = \frac{X_{\rm{gr,vert}}}{\cos\theta}-X_{\rm{max}},
\end{equation}
where $X_{\rm{gr,vert}}$ is the vertical slant depth of the ground\footnote{$X_{\rm{gr,vert}} = 870$ g/cm$^2$ for the Pierre Auger Observatory}, $\theta$ is the zenith angle of the shower axis and \xmax is the depth in which the shower reaches its maximum. For each simulated event, the $\theta$ corresponds to the true zenith angle as set in the input of the simulation and \xmax was taken directly from the Gaisser-Hillas function fitted to the longitudinal energy deposit profile. 

\section{Characterization of the muon ground energy spectrum}
\label{sec:carac}

In this section, the simulations described in~\cref{sec:sim} are used to characterize the energy spectrum of muons at ground level and to study its relations with primary mass and hadronic interaction models.

\begin{figure*}
  \centering
  \includegraphics[width=0.9\textwidth]{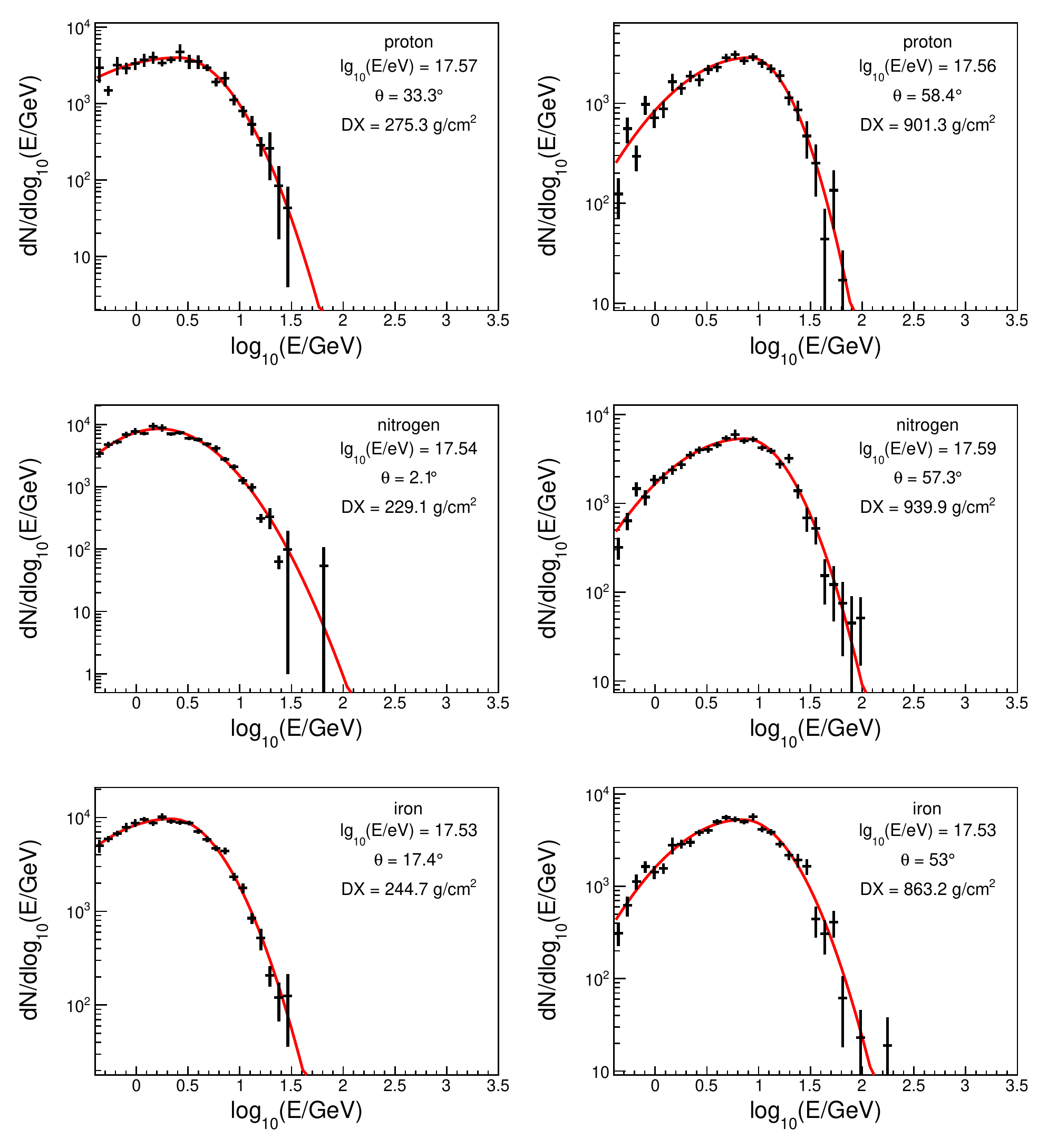}
  \caption{Example of the ground muon energy spectrum for six simulated showers. Shower parameters are shown in each panel. The energy spectra were built by collecting from the air-shower simulation all muons hitting the ground in a lateral distance between $425$ and $475$ meters from the shower axis. The low and high energy hadronic interaction models are FLUKA and QGSJetII-04, respectively. Red lines are the result of fitting~\cref{eq:carac:gaus} to the data points.}
  \label{fig:carac:examples}
\end{figure*}

\cref{fig:carac:examples} shows examples of the ground energy spectrum of muons for six simulated events, which differ by the primary particle and the shower geometry. The low and high energy hadronic interaction models are \fluka and \qgsjet, respectively. Examples have been selected to illustrate the general shape of the muon spectrum for different primary particles and extreme values of \dx. The left-hand column shows deep showers, with relatively small \dx values, while the right-hand column shows shallow showers, with relatively high values of \dx. The normalization and the mean of the distributions are clearly different but the overall shape is very similar.

As shown by the red lines in \cref{fig:carac:examples}, the ground energy spectrum of muons is well described by the following asymmetric Gaussian function

\begin{equation}
  \frac{\textrm{d}N_{\mu}}{\textrm{d}x} = \begin{cases}
    N_0 \exp\left[ -\frac{1}{2}\left(\frac{x-\eta}{\sigma}\right)^2 \right] , & \text{if $x<\eta$} \\
    & \\
    N_0 \exp \left[ -\frac{1}{2}\left(\frac{x-\eta}{\alpha\sigma}\right)^2 \right] , & \text{if $x>\eta$}
    \end{cases}
  \label{eq:carac:gaus}
\end{equation}
where $x = \log_{10}(E/\rm{GeV})$. $N_0$ is the normalization parameter and it is correlated with the total number of muons in the shower. $\eta$ is the mode of the energy distribution, and it is strongly correlated with the average energy of muons reaching ground between $425$ and $475$ meters distance from the shower axis. $\sigma$ and $\alpha$ give the width of the distribution, $\alpha$ being the parameter that measures the degree of asymmetry of the distribution.

Muon energy spectra of all simulated showers were fit by the function presented in~\cref{eq:carac:gaus}, in which $N_0$, $\eta$, $\sigma$ and $\alpha$ were taken as free parameters. The fitting was performed using a binned maximum likelihood method with Poissonian probability distribution functions.

\begin{figure}
  \centering

    \subfloat[]{\includegraphics[width=0.43\textwidth]{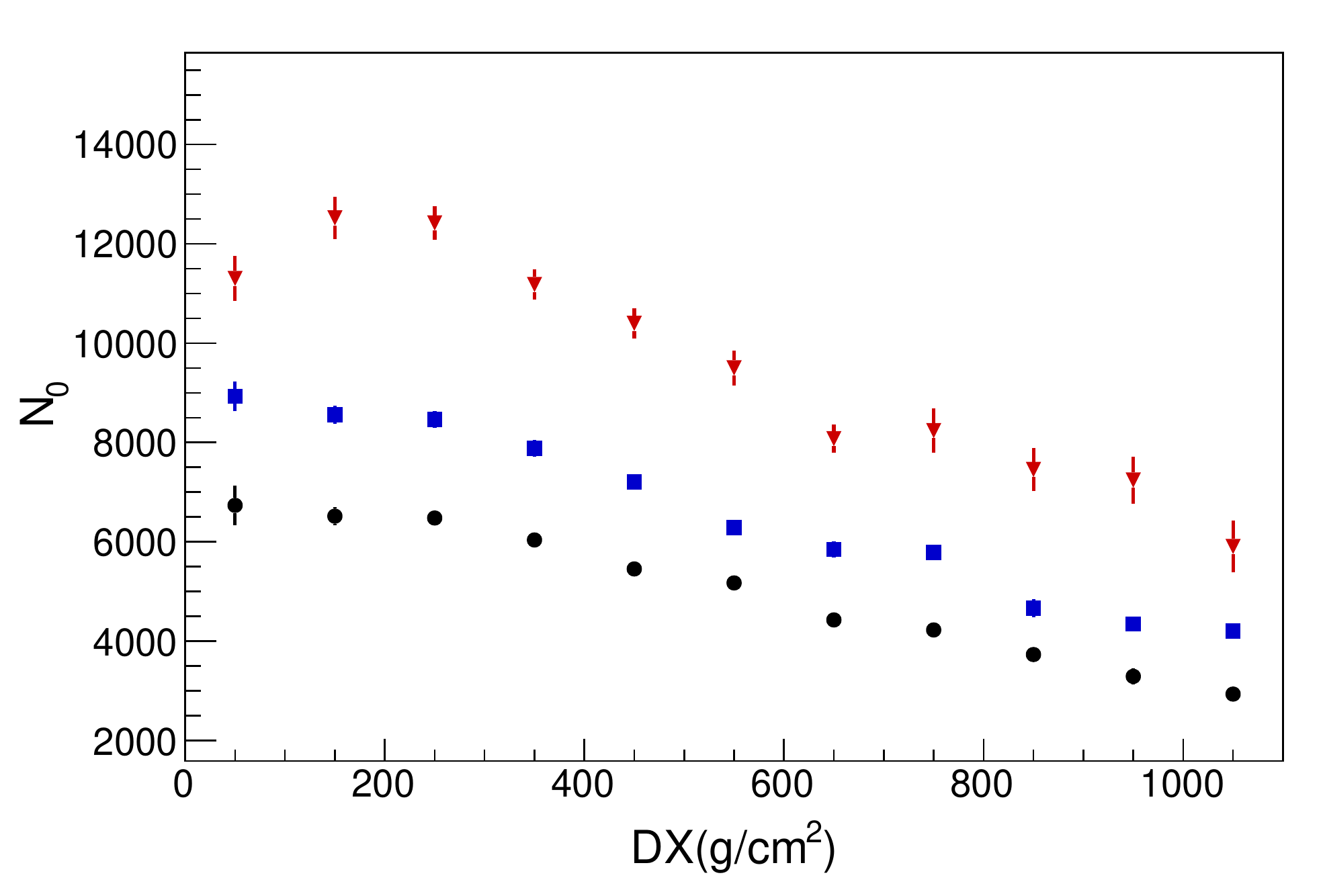}}
    \subfloat[]{\includegraphics[width=0.43\textwidth]{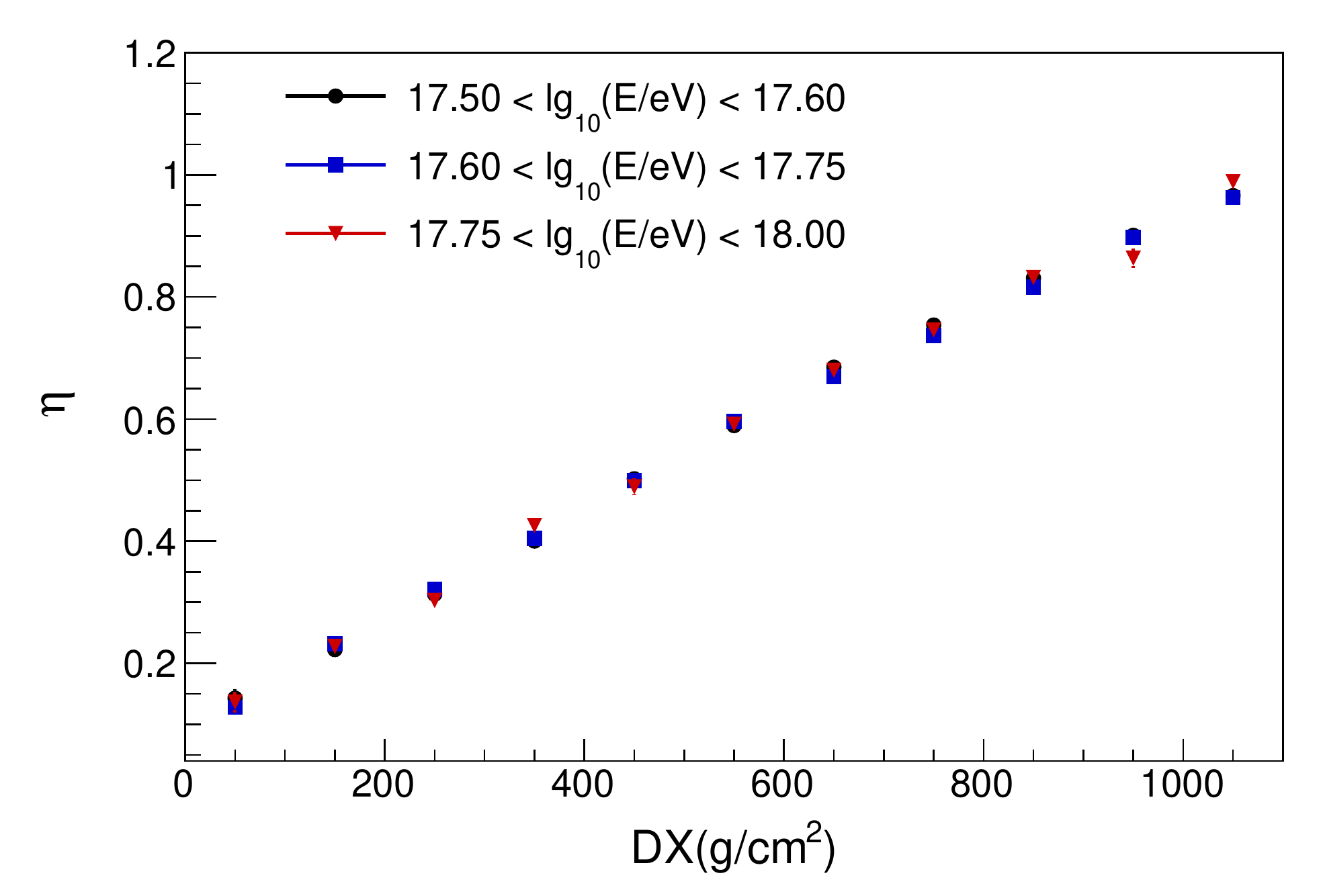}}

    \subfloat[]{\includegraphics[width=0.43\textwidth]{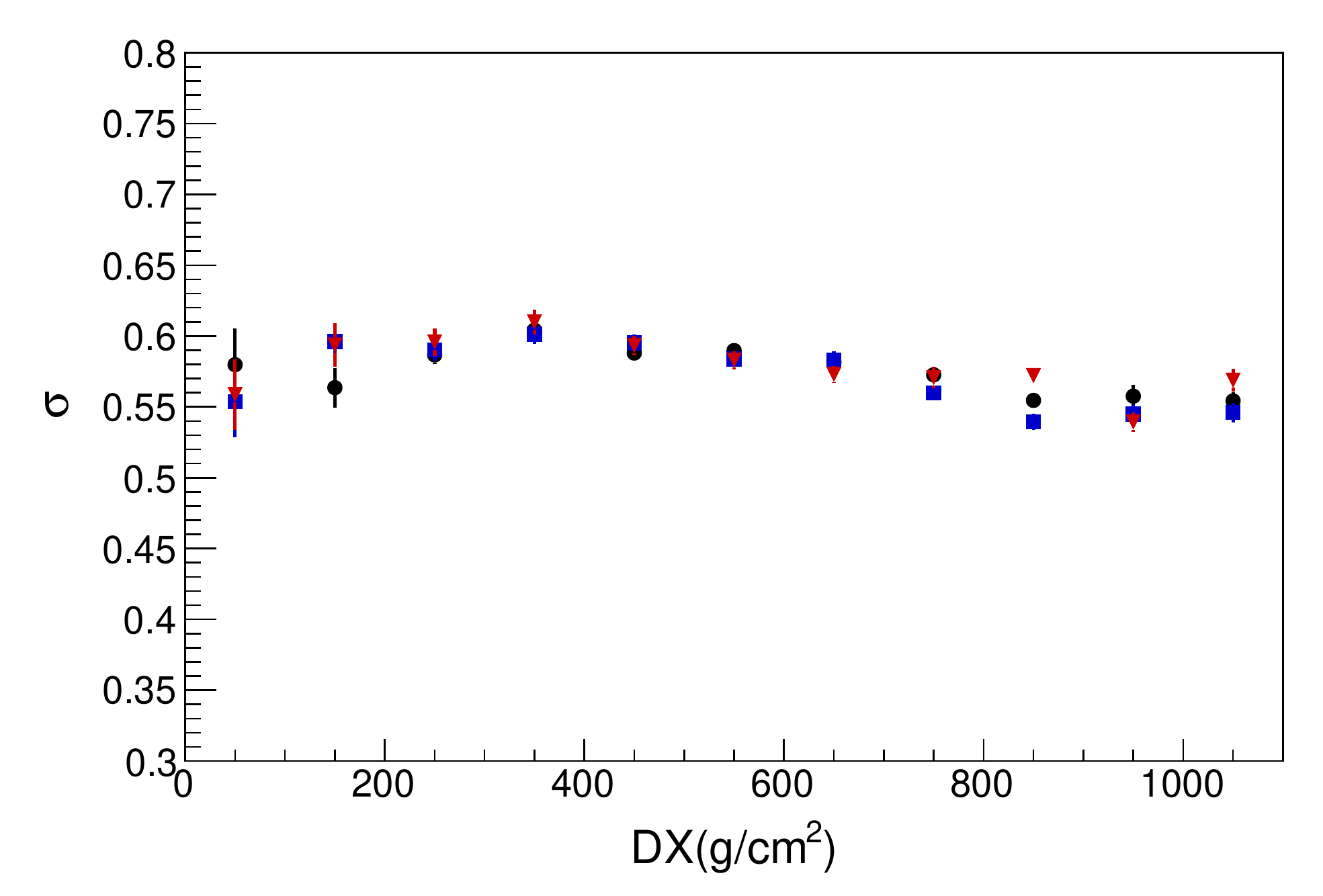}}
    \subfloat[]{\includegraphics[width=0.43\textwidth]{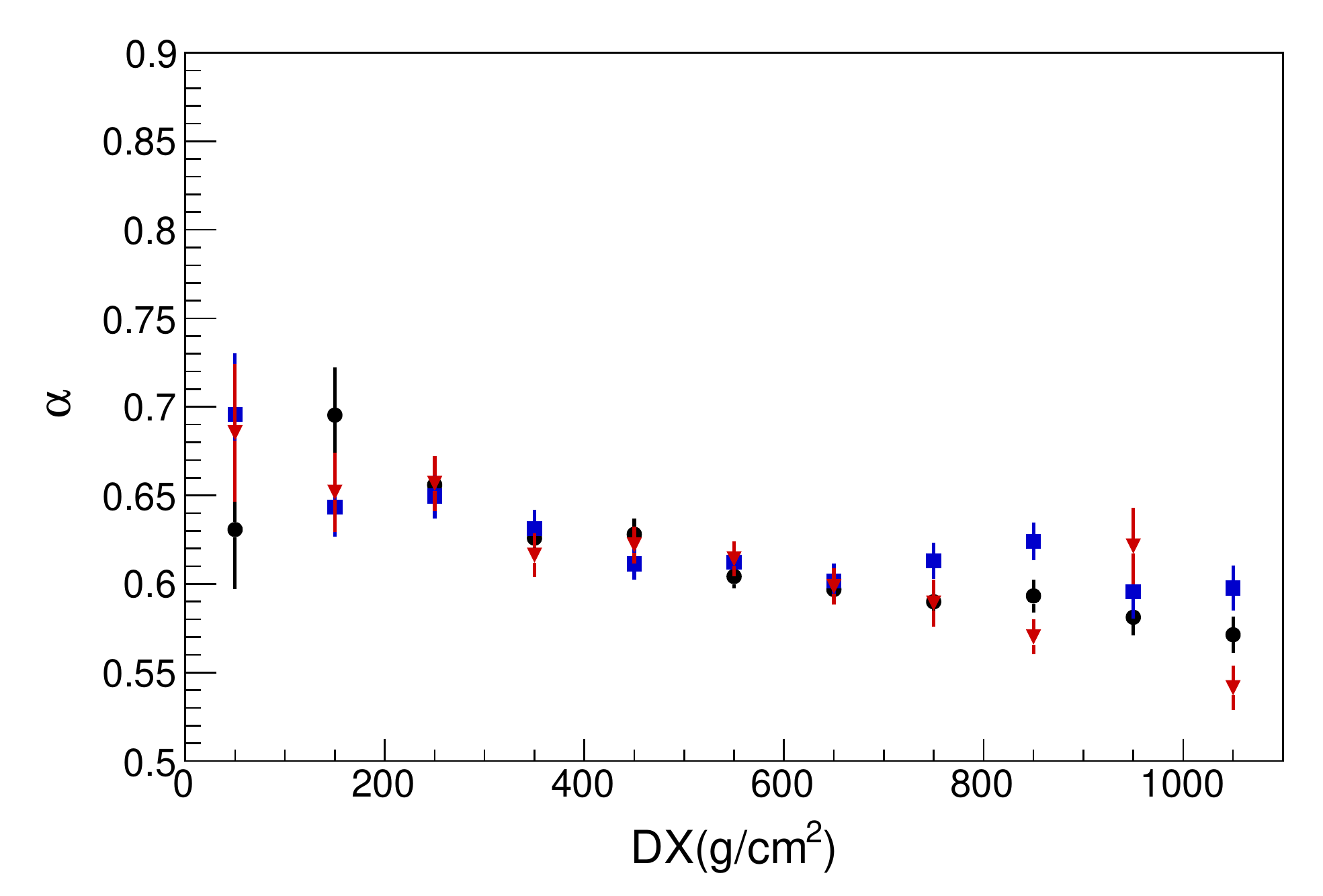}}

    \caption{(a) $N_0$, (b) $\eta$, (c) $\sigma$ and (d) $\alpha$ as a function of \dx for three energy intervals. Same number of p, N and Fe showers are considered. The hadronic interaction models used are \fluka and \qgsjet.}
  \label{fig:carac:en}
\end{figure}

\begin{figure}
  \centering

    \subfloat[]{\includegraphics[width=0.43\textwidth]{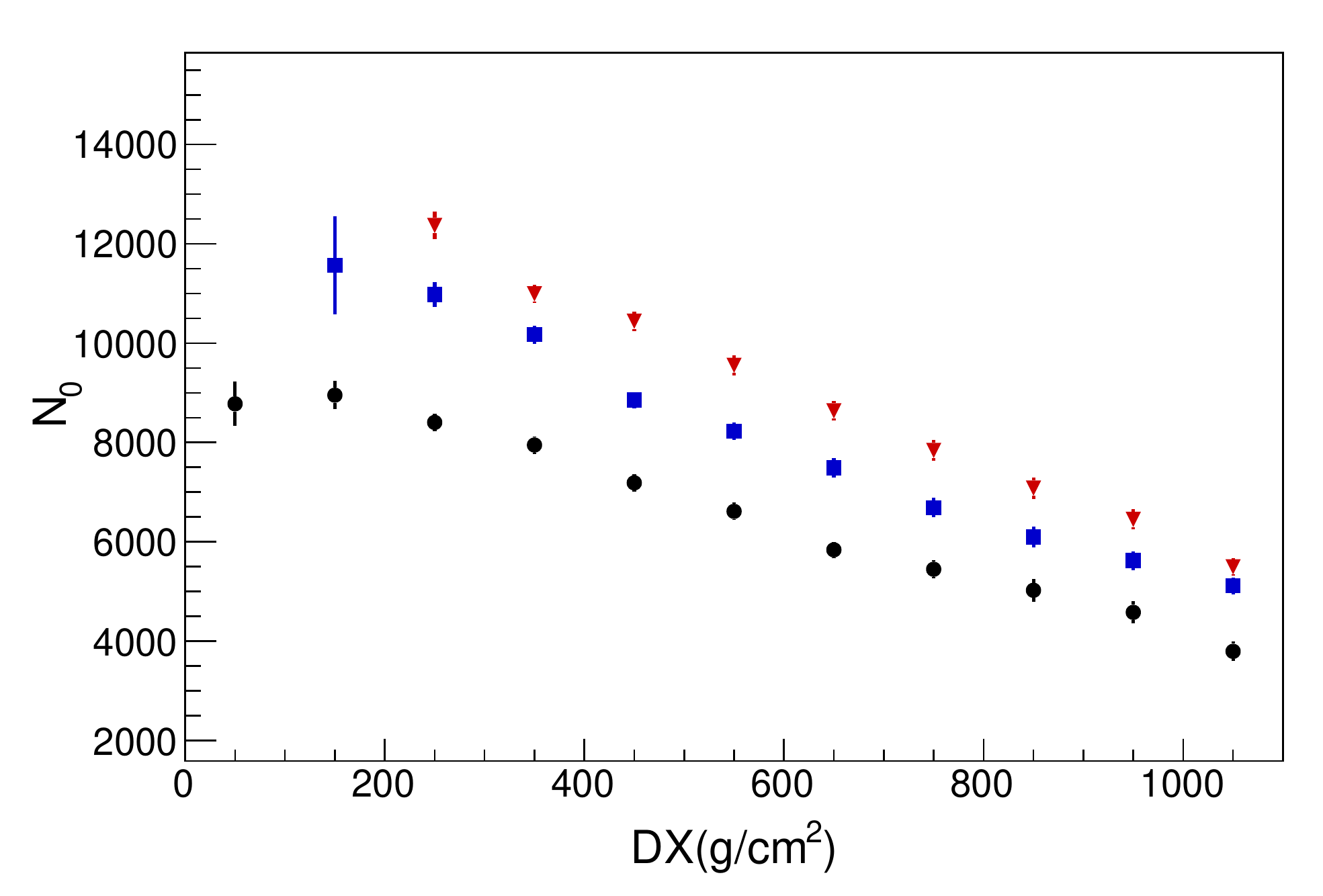}}
    \subfloat[]{\includegraphics[width=0.43\textwidth]{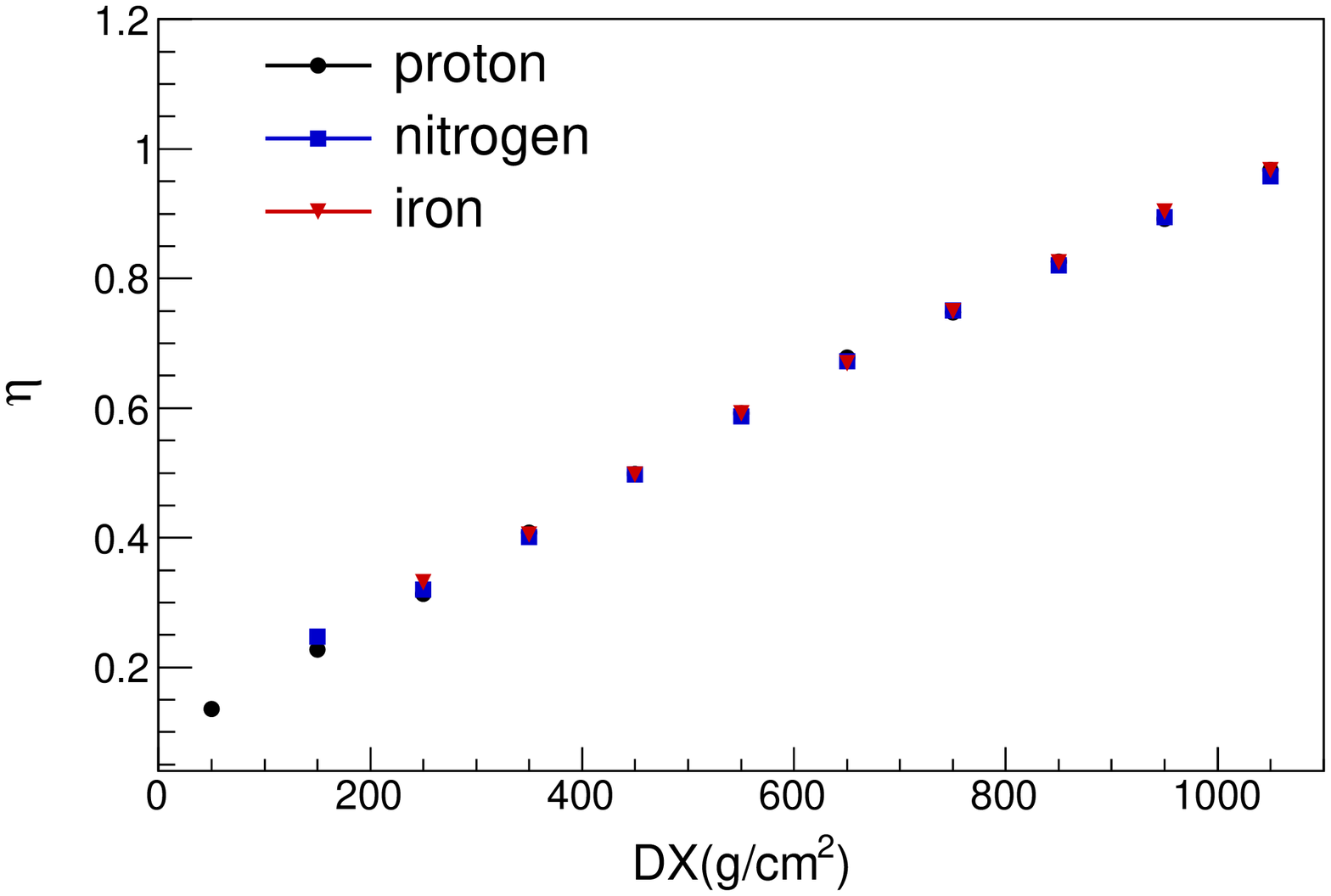}}

    \subfloat[]{\includegraphics[width=0.43\textwidth]{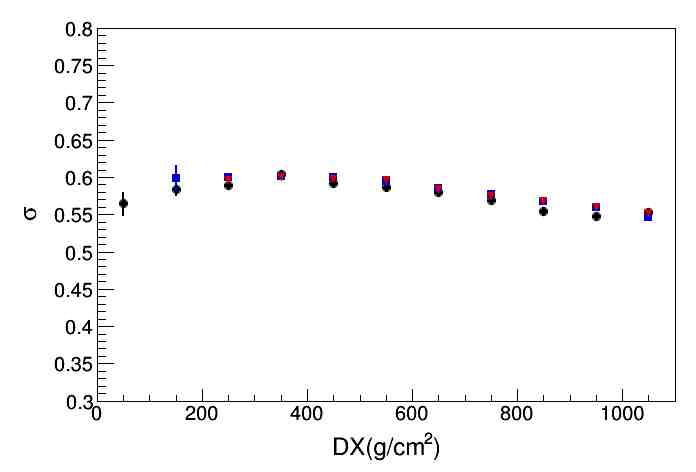}}
    \subfloat[]{\includegraphics[width=0.43\textwidth]{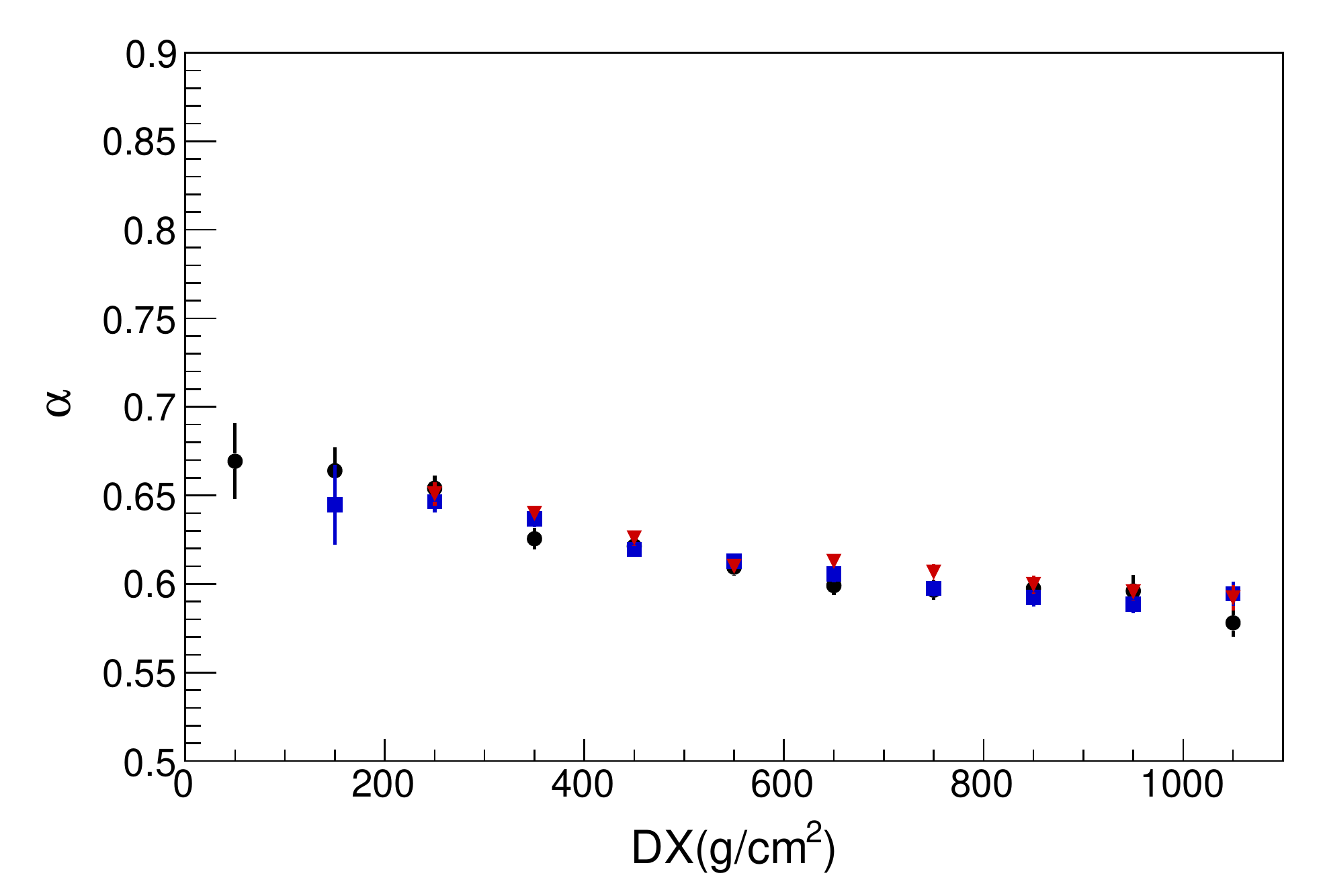}}
    \caption{(a) $N_0$, (b) $\eta$, (c) $\sigma$ and (d) $\alpha$ as a function of \dx for three primary particles. The energy interval for all primaries is $17.5 < \log_{10} (E/eV) < 18.0$. The hadronic interaction models used are \fluka and \qgsjet.}
    \label{fig:carac:mass}
\end{figure}

\begin{figure}
  \centering

    \subfloat[]{\includegraphics[width=0.43\textwidth]{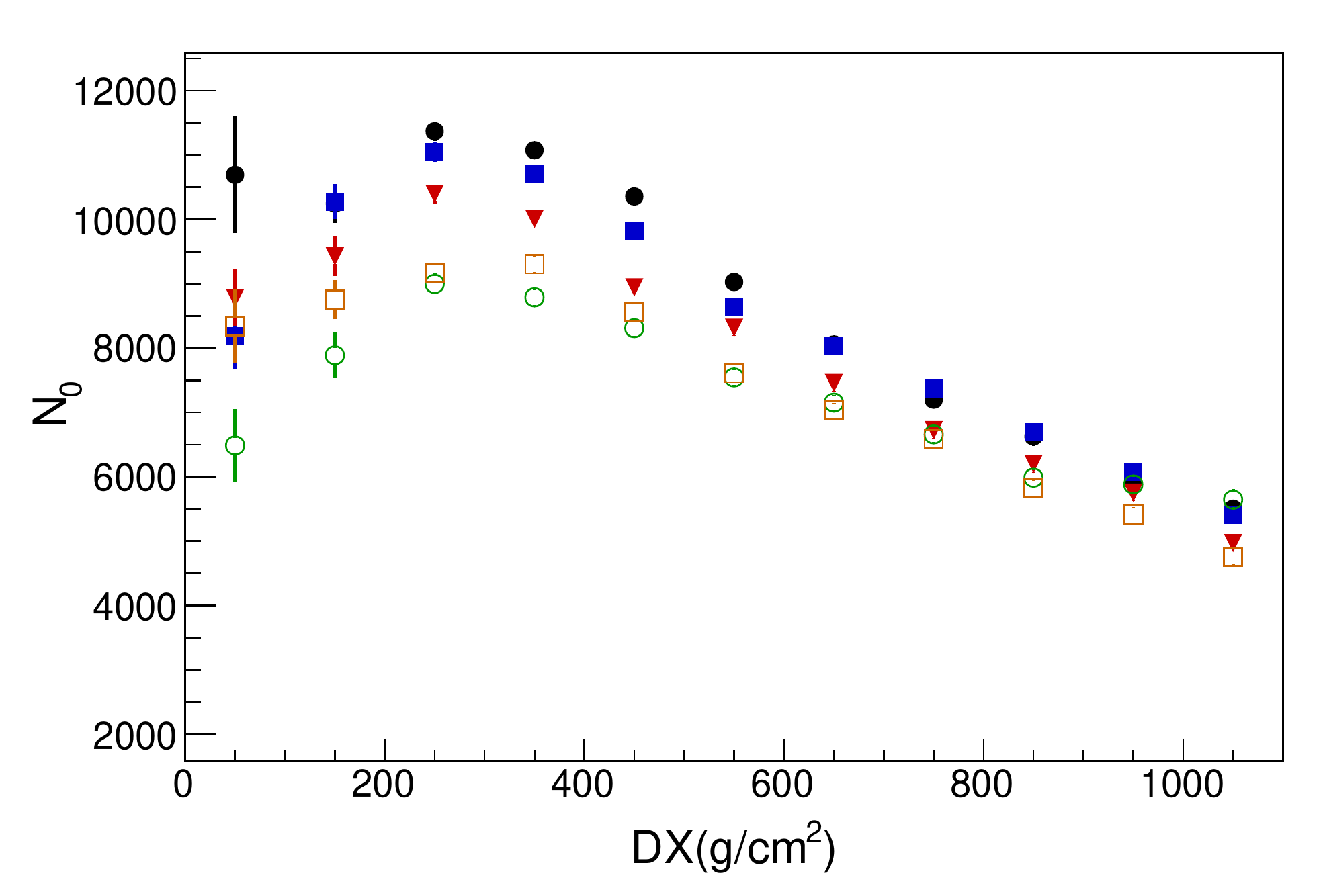}}
    \subfloat[]{\includegraphics[width=0.43\textwidth]{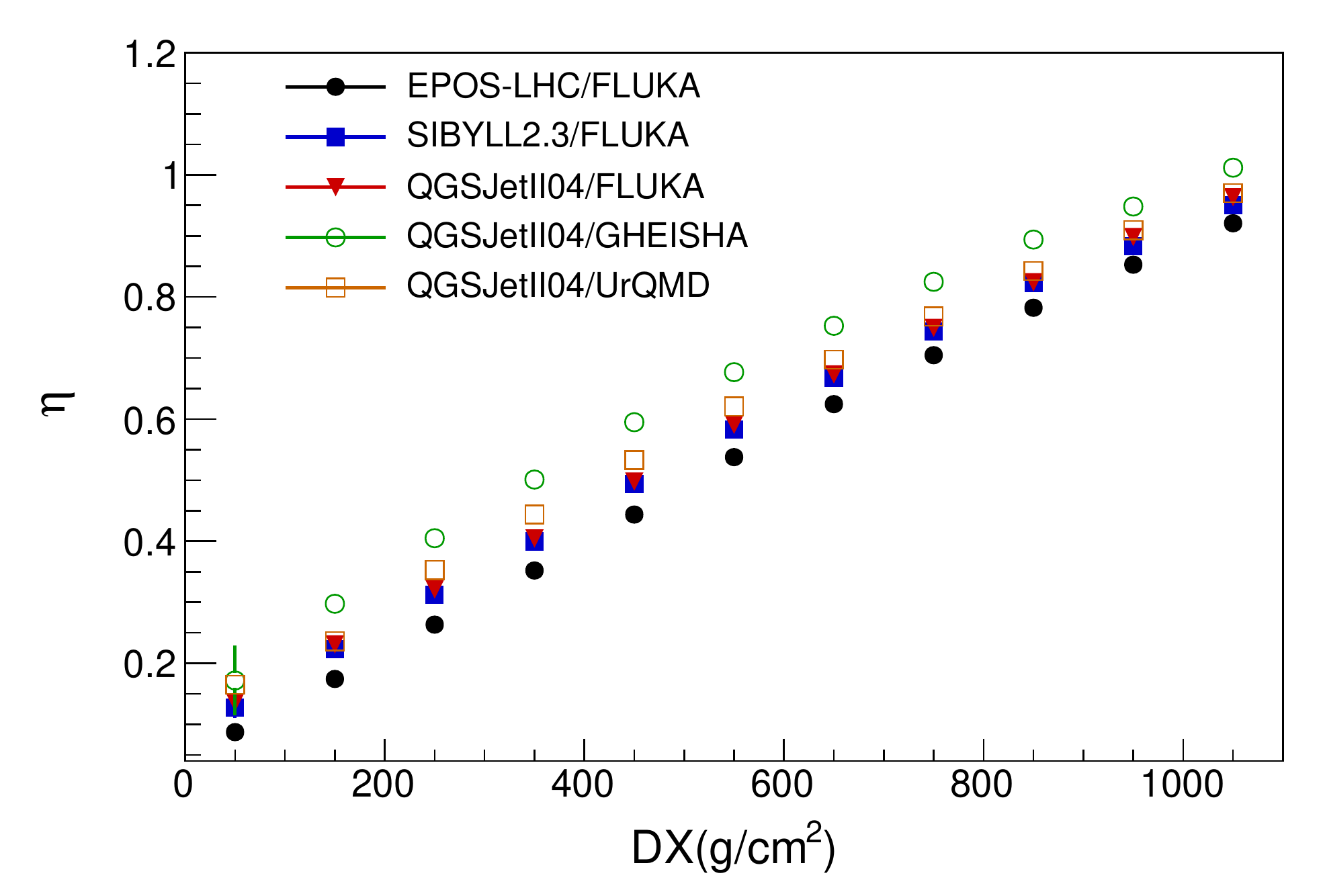}}

    \subfloat[]{\includegraphics[width=0.43\textwidth]{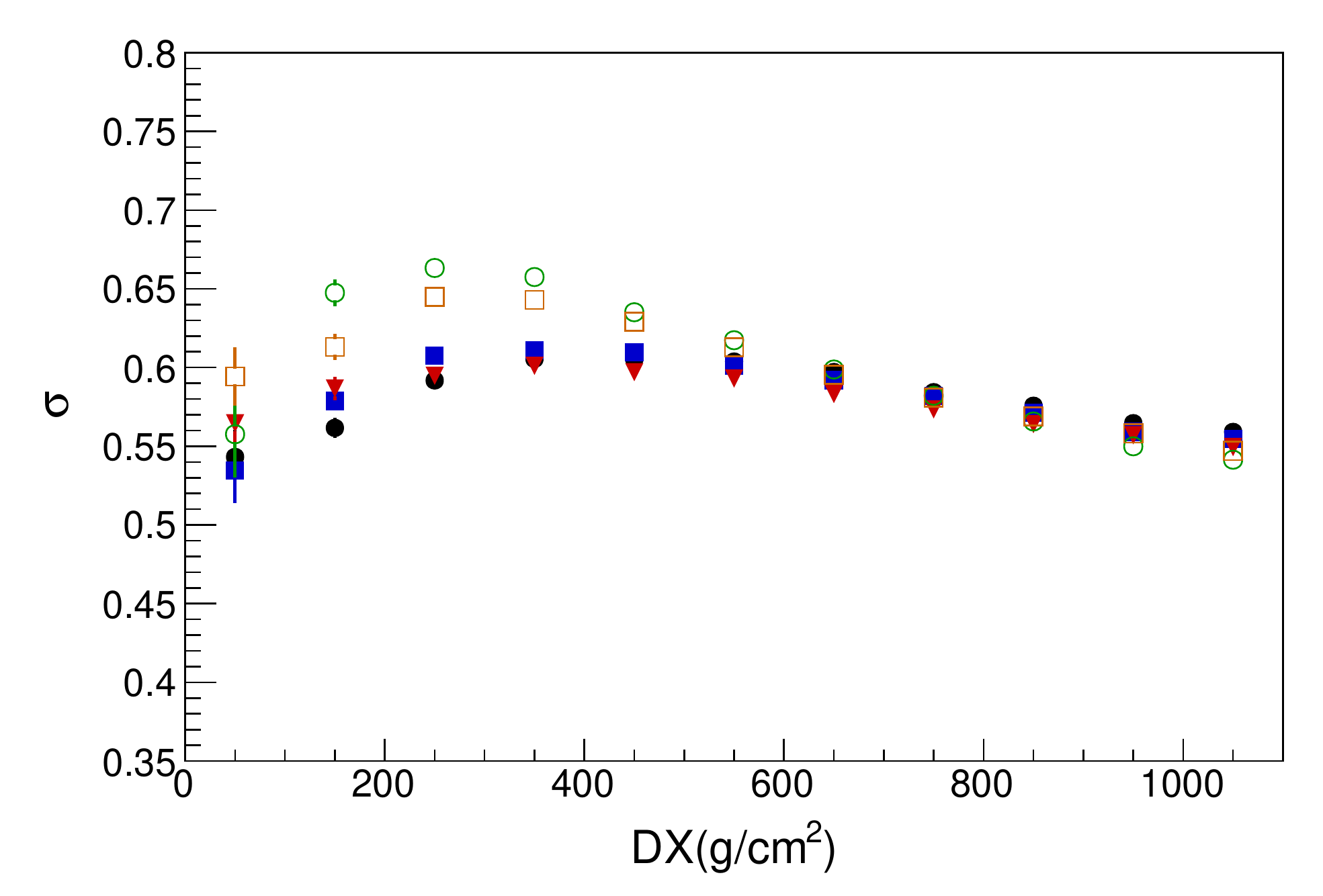}}
    \subfloat[]{\includegraphics[width=0.43\textwidth]{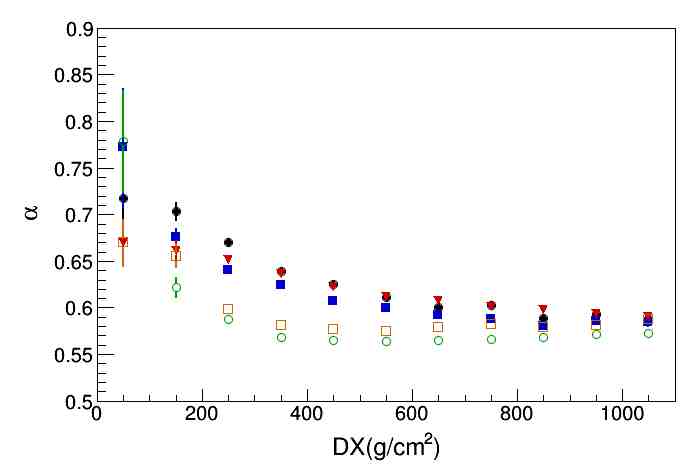}}

    \caption{(a) $N_0$, (b) $\eta$, (c) $\sigma$ and (d) $\alpha$ as a function of \dx for five combinations of hadronic interaction models. The energy interval for all primaries is $17.5 < \log_{10} (E/eV) < 18.0$. Same number of p, N and Fe showers are considered.}
  \label{fig:carac:mod}
\end{figure}

Since the ground energy spectrum of muons is well described by the asymmetric Gaussian function, one may study its dependencies on the energy, primary particle and hadronic interaction models through the evolution of the parameters $N_0$, $\eta$, $\sigma$ and $\alpha$ with \dx. \cref{fig:carac:en} shows the \dx evolution of the four parameters for three different energy intervals, \cref{fig:carac:mass} for three primary particles and \cref{fig:carac:mod} for five combinations of high and low energy hadronic interaction models.

It is clear from~\cref{fig:carac:en} that the normalization ($N_0$) is the only property of ground muon energy spectrum which shows a significant dependence on the primary energy. As expected, $N_0$ also depends strongly on the primary particle and hadronic interaction model (see~\cref{fig:carac:mass,fig:carac:mod}), which reflects the very known behavior of number of muons in air showers. Concerning the peak position of the distributions, $\eta$, a strong evolution with \dx is observed, revealing the shift of the average energy of muons to higher values with increasing \dx. Besides normalization and peak positions, changes on the overall shape of the energy spectrum of muons can be evaluated through the parameters $\sigma$ and $\alpha$. These parameters clearly show a very weak evolution with \dx, and a nearly null dependence on the primary energy and primary particle and a relatively very small dependence on the hadronic interaction models.

The analysis of~\cref{fig:carac:mod} also repeats known though less popular lessons: the effect of low energy hadronic interaction models are as important as the high energy one regarding the description of the muonic component in air-shower simulations~\cite{Meurer2006}. The differences of $N_0$ between \qgsjet/\fluka and \qgsjet/\urqmd or \gheisha are of the same order or larger than the largest difference between the high energy interaction models. The average energy of muons, correlated with $\eta$, is larger for \gheisha, followed by \urqmd and then \fluka. The differences in $\eta$ due to the low energy hadronic interaction models are as large as the differences due to the high energy hadronic interaction models. Regarding $\sigma$ and $\alpha$ parameters, one can see in~\cref{fig:carac:mod} they show the same weak \dx evolution for all the hadronic model combinations. Furthermore, the effect of low energy hadronic models is again clear as shown by the difference between \fluka and \gheisha, or \urqmd.

Finally, the analysis of~\cref{fig:carac:mod} also opens new possibilities. It turns out that $\eta$ is strongly sensitive to the hadronic interaction models, and at the same time, it does not show any significant dependence on the primary energy and on the primary particle. The lack of primary energy dependence is important experimentally to eliminate effects from the experimental energy scale, while the lack of primary particle dependence is an essential property to disentangle the hadronic interaction effects from the primary composition determination. The evolution of $\eta$ with \dx is simple (linear) and strong, which makes easy to study showers in different evolution stages.

All of these are indications that accessing experimentally the information carried by $\eta$ could be successfully used to constrain the hadronic interaction models. In the next section a new observable which is strongly correlated to $\eta$ is studied and its constraining power in realistic experimental conditions is tested.

\section{An observable to test hadronic interaction models}
\label{sec:obser}

Accessing $\eta$ directly is not possible for any running or planned UHECR experiment. Therefore, instead of proposing an unrealistic parameter, this study starts from a realistic experimental scenario and aims to find an observable which correlates to $\eta$. A generic experimental set-up is considered with two muon detector arrays with different amount of shielding leading to two energy thresholds. With such an experimental set-up, it is possible to measure the integral of the energy spectrum or the energy density of muons above two energy thresholds.

\begin{figure}
  \centering
  \includegraphics[width=0.7\textwidth]{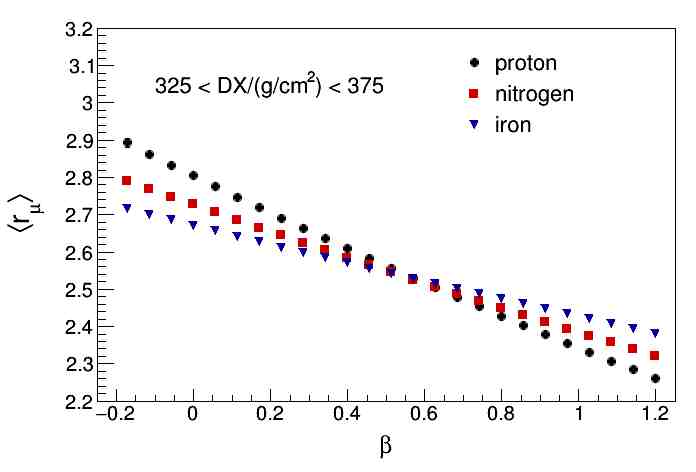}
  \caption{\rmumean as a function of $\beta$ for one small \dx interval and for three different primaries. $\beta = 0.6$ was chosen to minimize the dependence of \rmu with primary particle.}
  \label{fig:beta}
\end{figure}

Using the simulations described in~\cref{sec:sim} the density of muons in the lateral distance from 425 to 475 m from the shower axis was calculated for a surface (\smutop) and a buried (\smubur) detectors and a new parameter is defined:

\begin{equation}
  r_{\mu} = \sec^{\beta}\theta \; \frac{S_{\mu}^{\rm{bur}}}{S_{\mu}^{\rm{sur}}},
  \label{eq:rmu}
\end{equation}
where $\theta$ is the zenith angle of the primary particle. The term $\sec^{\beta}\theta$ compensates the zenith dependence on the energy threshold and the effective area of the buried detectors. The surface detector (\smutop) is considered to have $10$ m$^2$ and the buried detector (\smubur) to have $30/\cos\theta$ m$^2$, which are motivated by water-Cherenkov stations and flat buried scintillators respectively. $\beta$ was determined to minimize the primary mass dependence of \rmu. In~\cref{fig:beta} the \rmu dependence on $\beta$ is shown for one hadronic interaction models, which justify the choice of $\beta=0.6$ as the value that minimize the difference between primaries. The detector features considered is described in the next section.

\begin{figure*}
  \centering
  \includegraphics[width=0.32\textwidth]{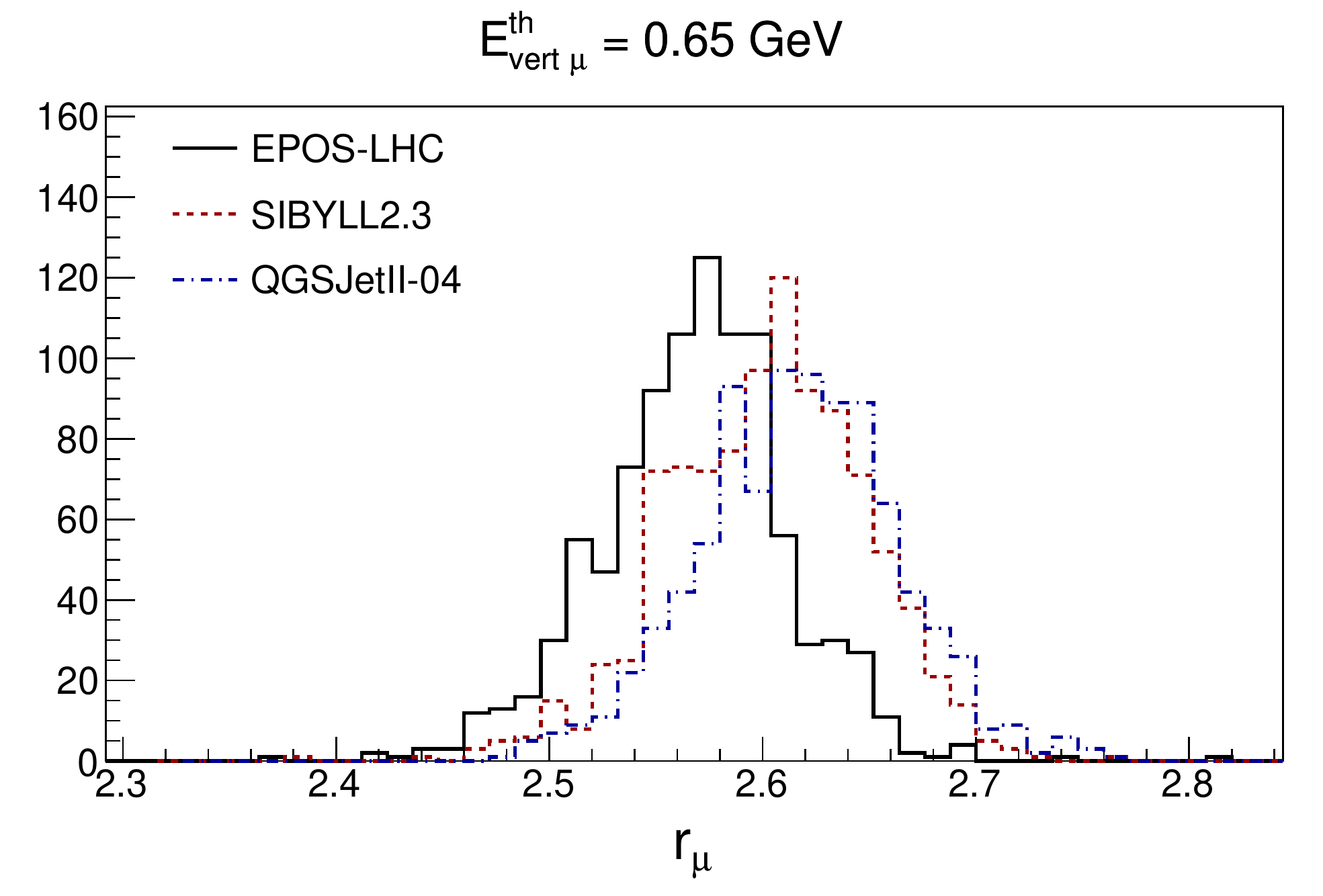}
  \includegraphics[width=0.32\textwidth]{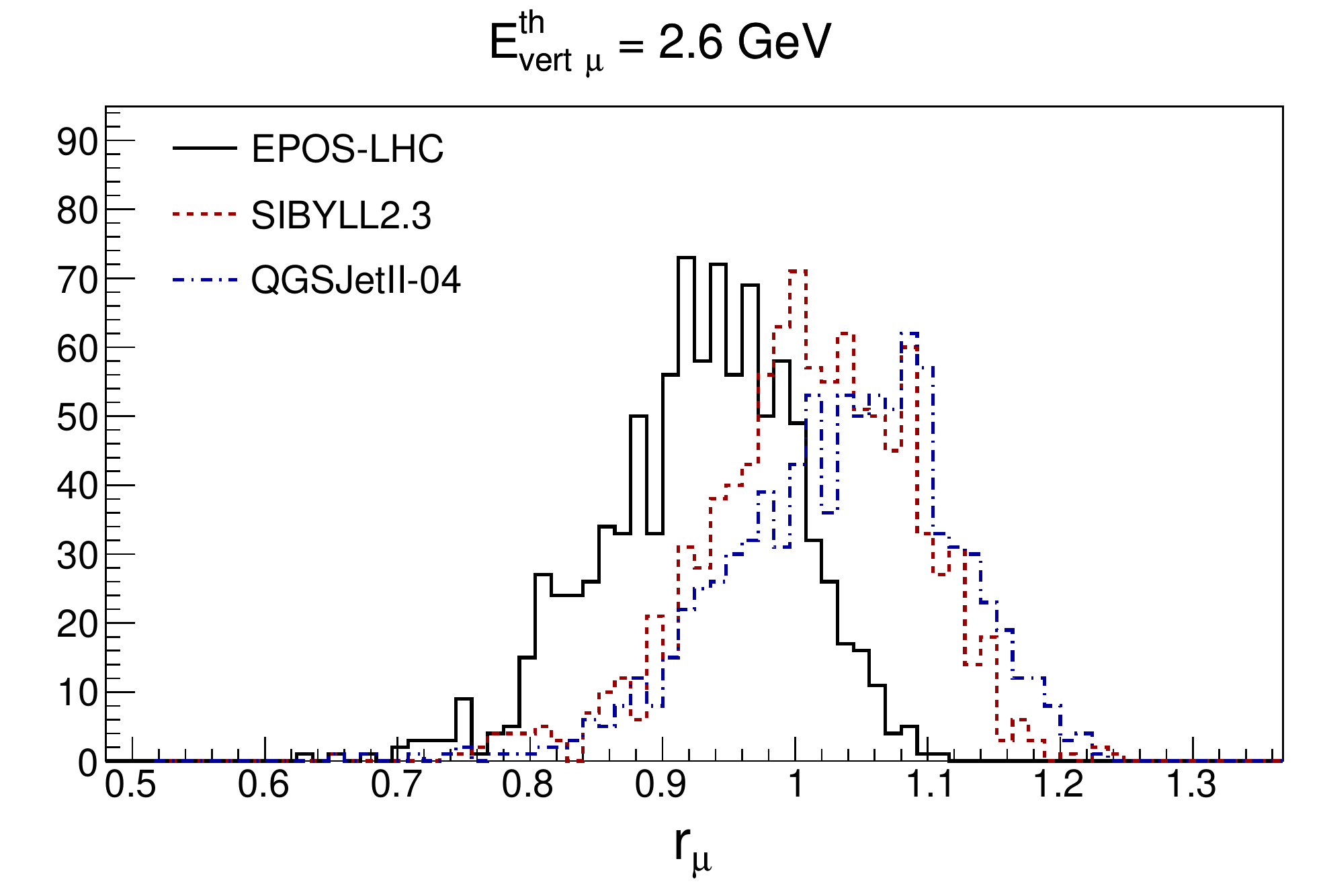}
  \includegraphics[width=0.32\textwidth]{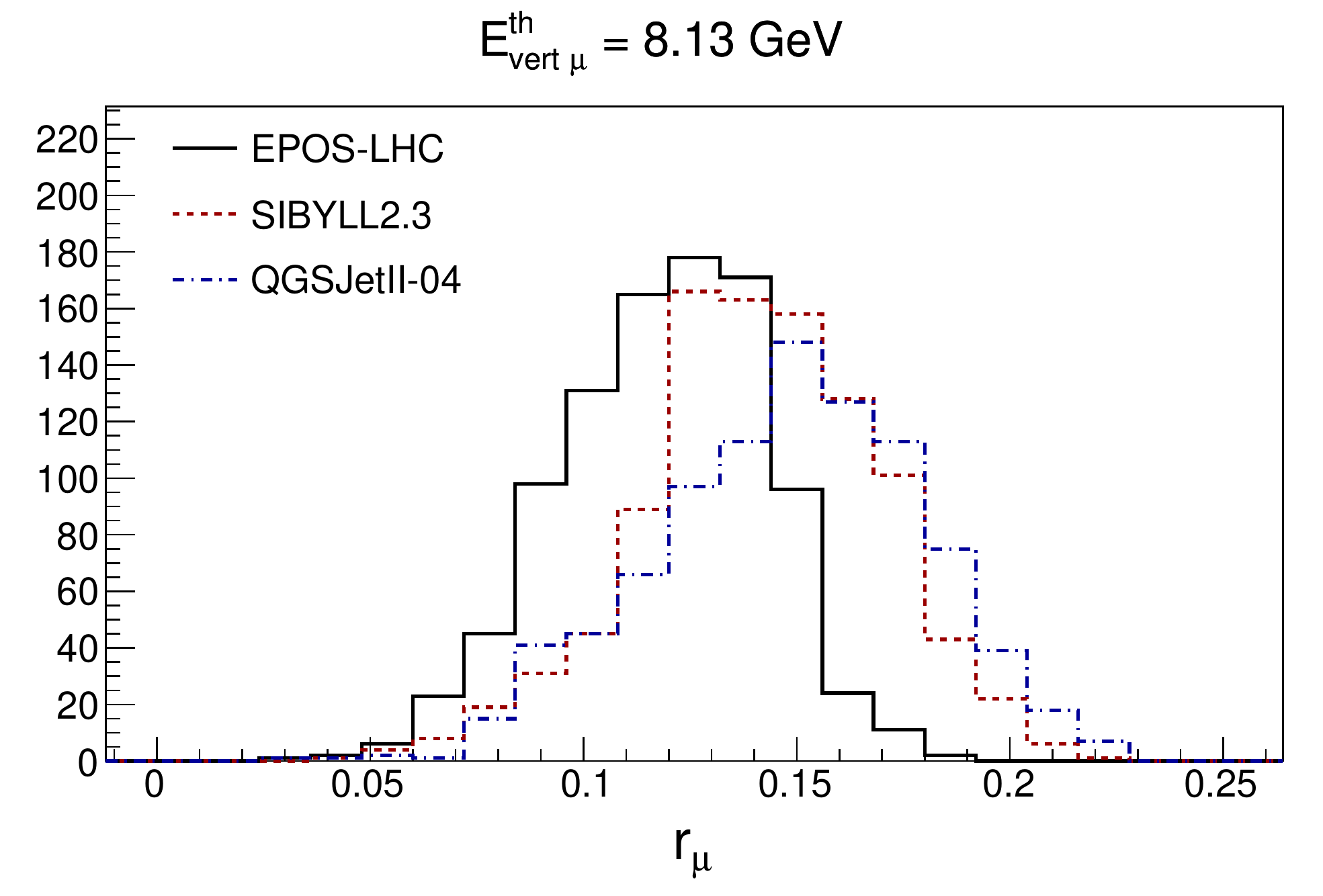}

  \caption{\rmu distributions of the three high energy hadronic interaction models. Each panel shows the distributions for a given energy thresholds of the buried detector, $E^{\textrm{th}}_{\textrm{vert}, \mu} = 0.65, \; 2.6, \; 8.13$ GeV from left to right. Low energy hadronic interaction model is \fluka.}
  \label{fig:distri:rmu:he}
\end{figure*}

\begin{figure*}
  \centering
  \includegraphics[width=0.32\textwidth]{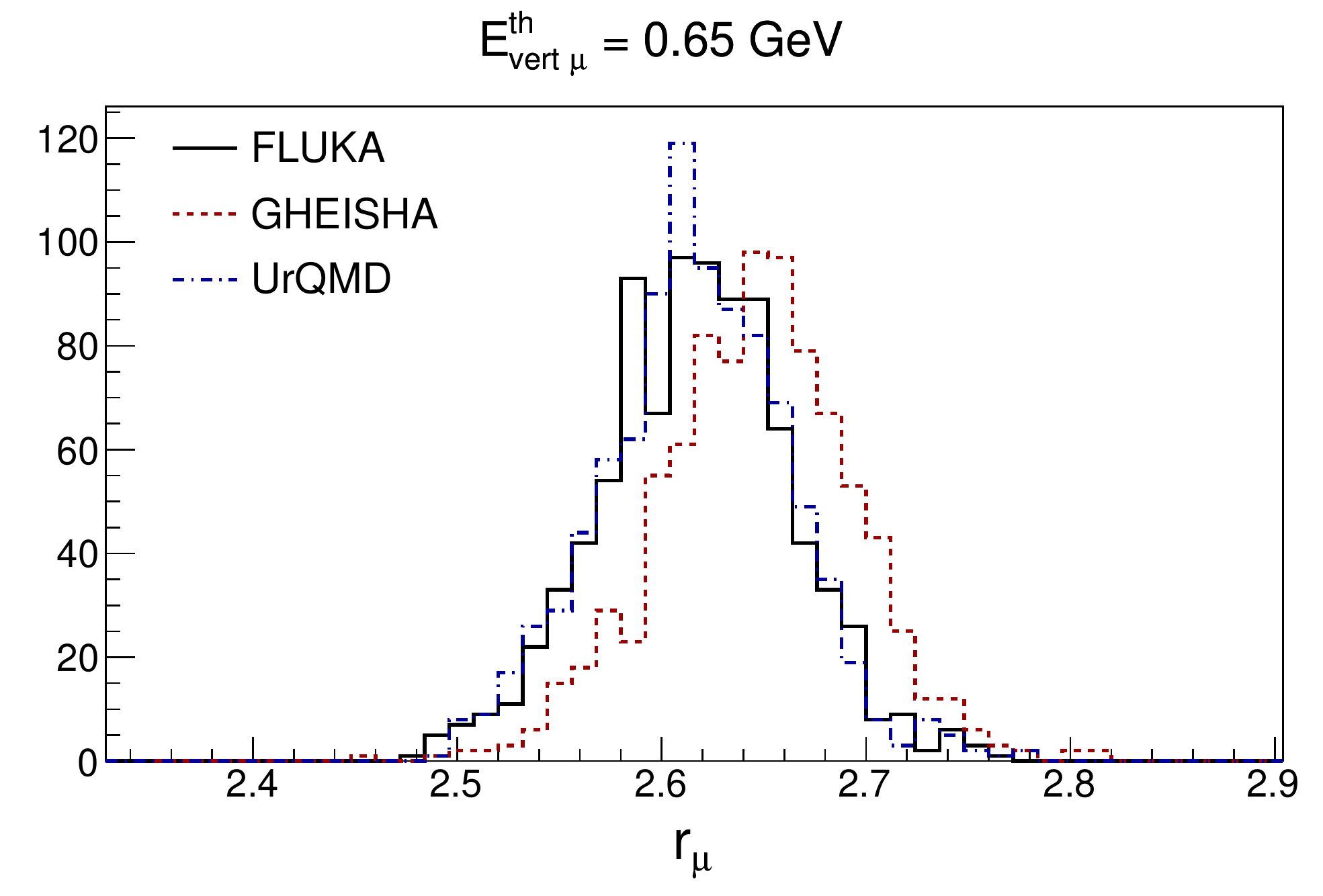}
  \includegraphics[width=0.32\textwidth]{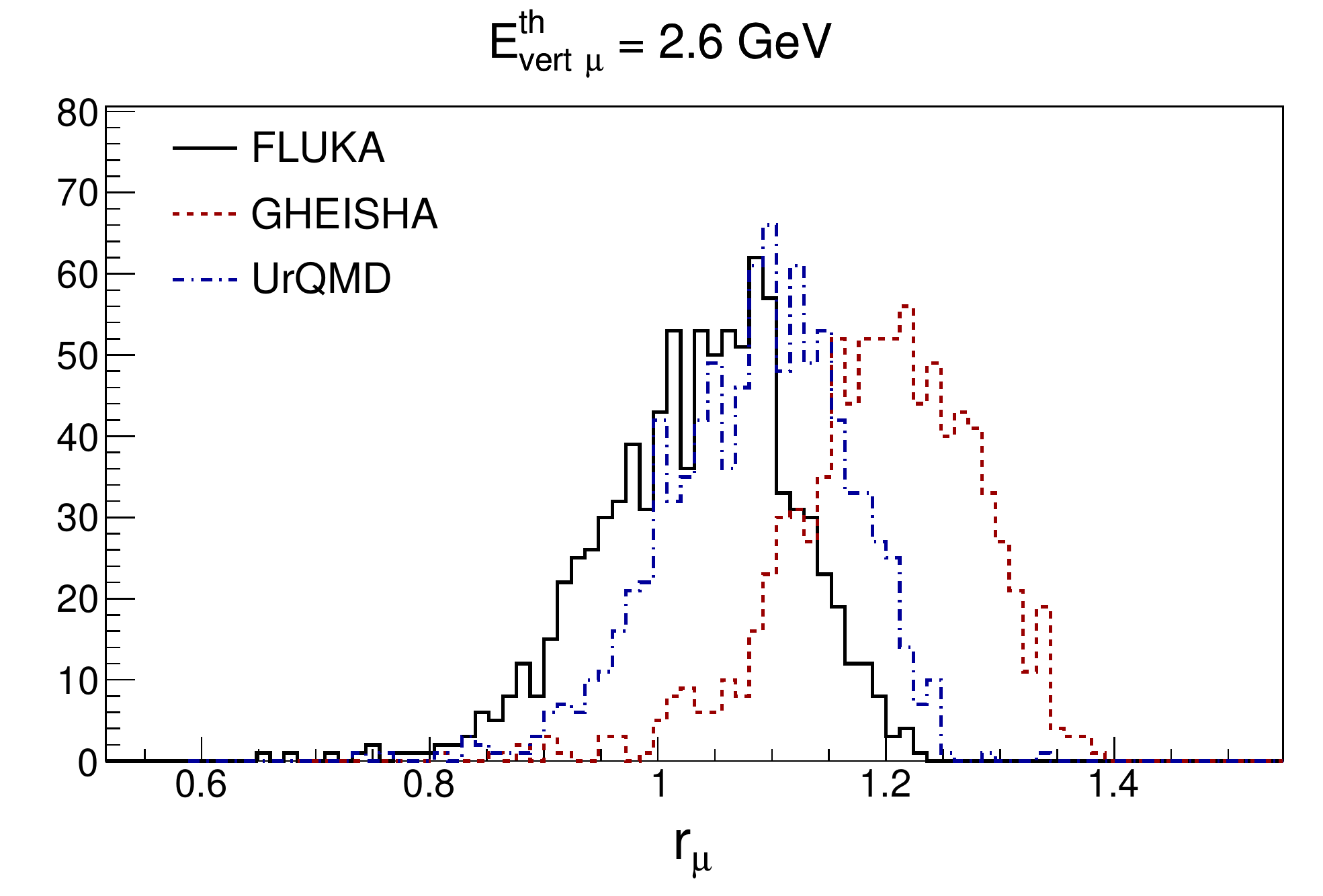}
  \includegraphics[width=0.32\textwidth]{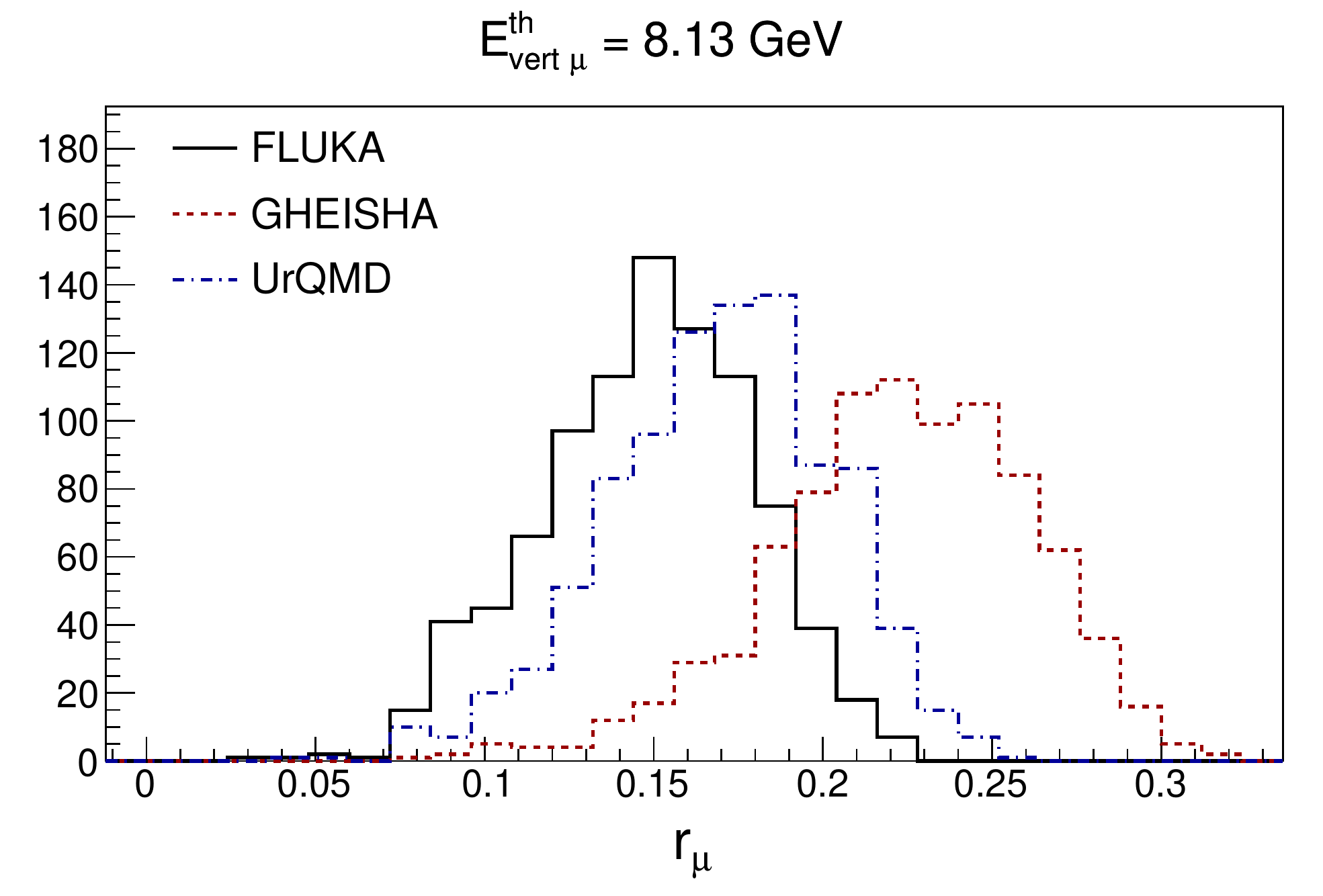}
   \caption{\rmu distributions of the three low energy hadronic interaction models. Each panel shows the distributions for a given energy thresholds of the buried detector, $E^{\textrm{th}}_{\textrm{vert}, \mu} = 0.65, \; 2.6, \; 8.13$ GeV from left to right. High energy hadronic interaction model is \qgsjet.}
  \label{fig:distri:rmu:le}
\end{figure*}

\cref{fig:distri:rmu:he,fig:distri:rmu:le} show the distributions of \rmu for several energy thresholds of the buried detector. Three cases are shown in which the vertical muon energy threshold of the buried detector ($E_{\rm{vert}, \mu}^{\rm{th}}$) is changed. The effective energy threshold for a muon with incident zenith angle $\theta_{\mu}$ is $E_{\rm{vert}, \mu}^{\rm{th}}/\cos\theta_{\mu}$. For the surface detectors, the energy threshold is kept fixed at $0.3$ GeV. \cref{fig:distri:rmu:he} compares the distributions for different high energy hadronic interaction models and \cref{fig:distri:rmu:le} for different low energy hadronic interaction models. The different degrees of separation of the hadronic interaction models with different energy threshold is clear, pointing to the possibility to constrain the hadronic interaction models using \rmu. In the next sections, the constraining power of \rmu is estimated and its correlation with $\eta$ is shown.

\subsection{\rmu determination including detector characteristics}
\label{sec:obser:detector}

In this section, the effect of detector geometry, muon energy thresholds, detector resolutions and systematic uncertainties on \rmu are studied. To include the detector features in the proposed analysis, the general characteristics of the muon detectors of the Pierre Auger Observatory are considered. A surface detector with $10$ m$^2$ of effective area is considered to measure muons with energies above $0.3$ GeV. Its general characteristics are inspired in the AugerPrime~\cite{Engel2015,AugerPDR,Gonzalez2016} design of water-Cherenkov stations with plastic scintillators on top. A second detector is considered based on the general characteristics of the Auger AMIGA detector~\cite{AMIGA_ICRC,AMIGA}. Those are $30$ m$^2$ flat scintillator detectors buried $2.5$ m below the ground. The muon energy threshold for the buried detectors depends on the incident zenith angle of the particle and it is given by $E_{\rm{th}} = \beta \rho h / \cos\theta_{\mu}$, where $\beta = 1.808$ MeV cm$^2$ g$^{-1}$ is the fractional energy loss per depth of standard rock, $\rho=1.8$ g cm$^{-3}$ is the soil density, $h=2.5$ m is the vertical depth of the detectors and $\theta_{\mu}$ is the zenith incidence angle of the muon. Because of its flatness, the effective collection area of the detector decrease by a factor $\cos\theta$, where $\theta$ is the zenith angle of the shower. The reconstruction of the muon density by an AMIGA-like detector has been shown to be satisfactorily possible for events with $\theta < 45^{\circ}$~\cite{Supanitsky2008,Ravignani2015,Ravignani2016}.

From now on, the results were obtained by considering the detector features shown above to calculate the muon density from the simulated air showers explained in~\cref{sec:sim}. In this way, the most important properties of the detectors are considered without the need to perform a full detector simulation.

\begin{figure}
  \centering
  \includegraphics[width=0.7\textwidth]{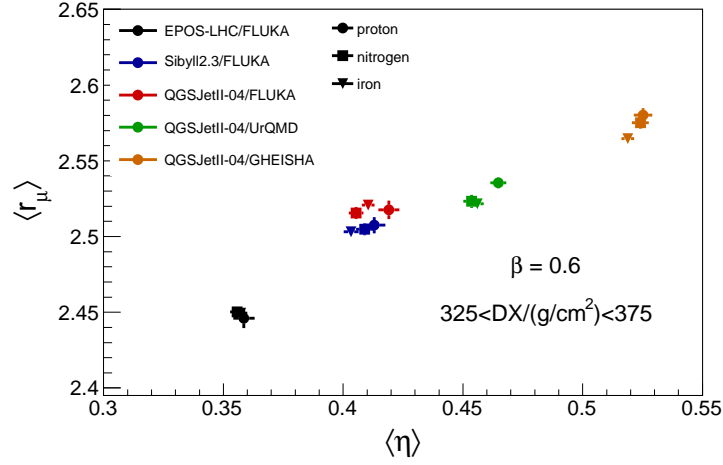}
  \caption{Mean \rmu as a function of $\langle \eta \rangle$ for $325 < DX/(\textrm{g/cm}^2)<375$. $\beta$ was set to 0.6. See~\cref{eq:carac:gaus,eq:rmu} for definitions of the parameters. Detectors effective area and threshold were considered according to~\cref{sec:obser:detector}. Five combinations of hadronic interaction models is shown for three primary particles. A nearly linear correlation is seen and a clear separation of many hadronic interaction models is visible.}
  \label{fig:obser:had}
\end{figure}

\cref{fig:obser:had} shows the relation between the average value of \rmu (\rmumean) and the average value of $\eta$ ($\langle \eta \rangle$) for showers with $325 < DX/(\rm{g/cm}^2)<375$ according to the detector properties explained above. Five combinations of low and high energy hadronic interaction models are shown in different colors and three primaries are shown in different marker styles. From~\cref{fig:obser:had}, one can see a nearly linear relation between \rmumean and $\langle \eta \rangle$. Furthermore, the relatively large separation between the hadronic interaction model combinations by \rmumean is clear, which shows \rmu is a good observable to constrain hadronic interaction properties.

\begin{figure}
  \centering
  \includegraphics[width=0.7\textwidth]{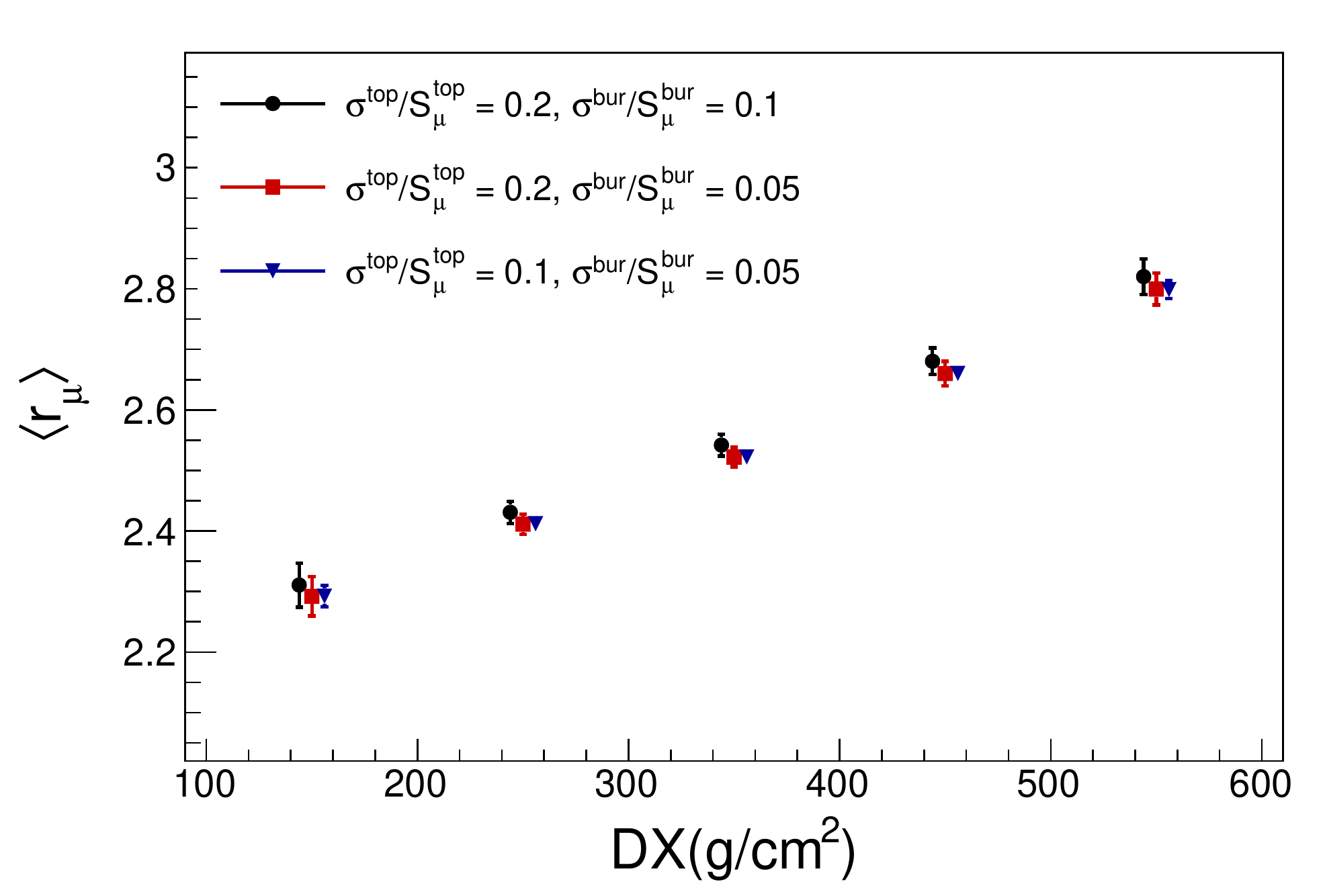}
  \caption{\rmumean as a function of \dx for different values of \smutop and \smubur resolution. The mean value \rmumean is calculated over 2000 realizations in which the detector resolution was applied to each simulated shower. Points were artificially shifted in \dx for clarity.}
  \label{fig:obser:dx:exp}
\end{figure}

The resolutions on the muon density reconstruction were taken into account by applying a Gaussian smearing on the true signal obtained from the simulations. \smutop resolution was considered to vary in the range from $10\%$ to $20\%$~\cite{Engel2015,Gonzalez2016} and \smubur resolution in the range from $5\%$ to $10\%$~\cite{Ravignani2015,Ravignani2016}. \cref{fig:obser:dx:exp} shows the effect of the resolution in the calculation of \rmumean. The Gaussian smearing were performed 2000 times and the \rmumean is shown as a function of \dx. The standard deviation of \rmu distributions is shown by the error bars. Three cases are shown in which the resolution on both \smutop and \smubur vary. The systematic effect of the detector resolution in the determination of \rmumean is smaller than 5\%.

\begin{figure}
  \centering
  \includegraphics[width=0.47\textwidth]{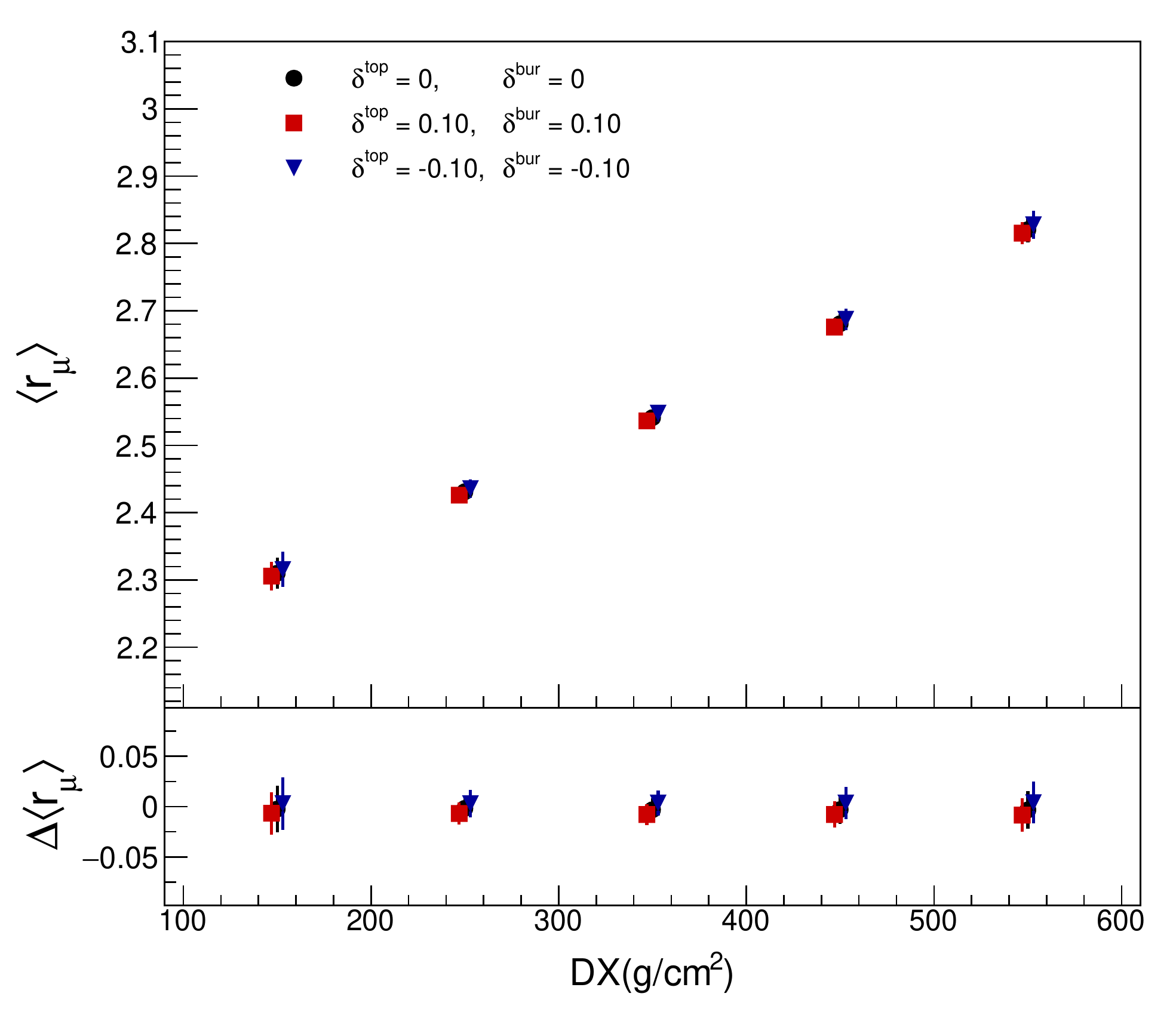}
  \includegraphics[width=0.47\textwidth]{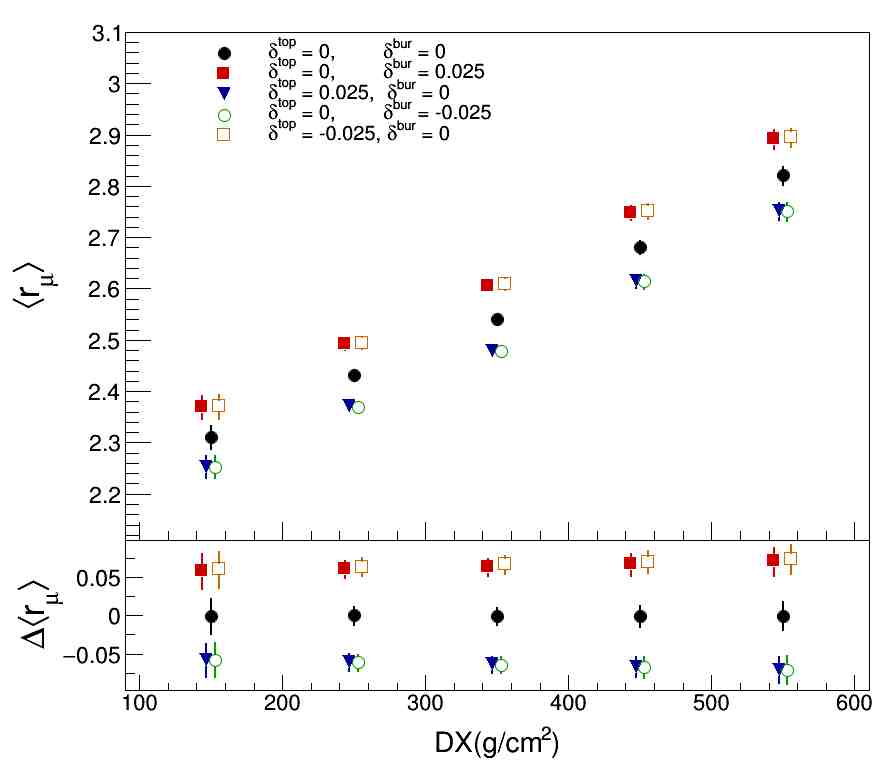}
  \caption{\rmumean as a function of \dx for different combinations of the systematic uncertainties on \smutop and \smubur. $\delta^{top}$ and $\delta^{bur}$ are defined in \cref{sec:obser:detector}. All simulated primary particles are included (p, N, Fe). Bottom panel shows the difference with relation to the average value. The hadronic interaction model combination used is \qgsjet/\fluka. Points were artificially shifted in \dx for clarity.}
  \label{fig:obser:dx:syst}
\end{figure}

Besides the experimental resolutions, systematic uncertainties on \smutop and \smubur can be originated from the detection and reconstruction procedures. Typically the most significant systematic uncertainty on muon density measurements are due to systematic uncertainties on the shower energy determination, which affects both \smutop and \smubur in the same magnitude. To evaluate the effect of systematic uncertainties on \rmu, the simulated \smutop and \smubur were shifted artificially and the resulting \rmu were calculated. First it was considered the shifts on \smutop and \smubur are totally correlated, which means the same magnitude and direction. This case represents the energy reconstruction uncertainty effect. \cref{fig:obser:dx:syst} shows the \rmu as a function of \dx, for one hadronic interaction models combination, for different cases in which \smutop and \smubur were shifted by a factor $1+\delta^{top}$ and $1+\delta^{bur}$ respectively. In~\cref{fig:obser:dx:syst} left panel it is shown the effect of a 10\% shift on both muon density at the same direction. Clearly, correlated systematic uncertainties on \smutop and \smubur have an insignificant effect on \rmu. In~\cref{fig:obser:dx:syst} right panel it is shown the effect of systematic shifts of 2.5\% on \smutop and \smubur in opposite directions. The magnitude of the \rmumean deviation is of order 0.05. The consequences of this deviation on the separation between different hadronic interaction models is discussed in~\cref{sec:results}.

\subsection{\rmu dependence on primary mass and energy}

For the following analysis, the \dx range was defined to preserve a good statistics in all \dx bins and it goes from $100$ to $600$ g/cm$^2$ divided in $5$ bins of $100$ g/cm$^2$. The upper bound of 600 g/cm$^2$ is highly influenced by the shower zenith angle limitation at $\theta < 45 ^{\circ}$ due to the buried detector features.

\begin{figure}
  \centering
    \includegraphics[width=0.7\textwidth]{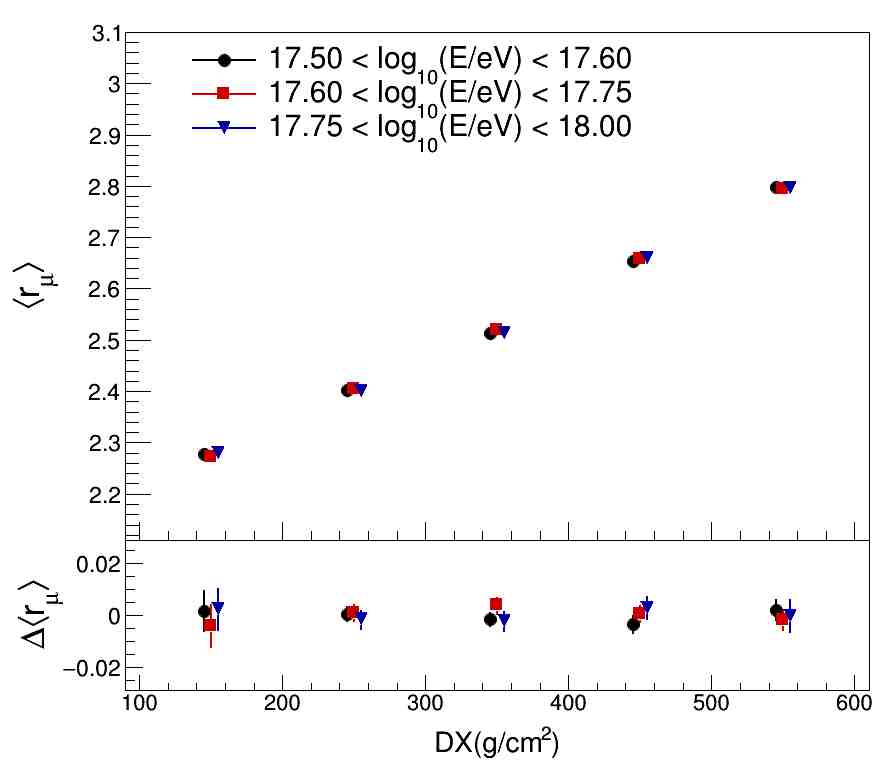}
    \caption{\rmumean as a function of \dx for three energy intervals.  All simulated primary particles are included (p, N, Fe). Bottom panel shows the difference with relation to the average value. The hadronic interaction model combination used is \qgsjet/\fluka. Points were artificially shifted in \dx for clarity.}
  \label{fig:obser:dx:en}
\end{figure}

\cref{fig:obser:dx:en} shows the evolution of \rmumean as a function of \dx for different energy ranges. All simulated primary particles are include (p, N and Fe). Hadronic interaction model combination here is \qgsjet/\fluka. A better visualization of the differences in \rmumean is seen in the bottom panel of~\cref{fig:obser:dx:en}, where \deltarmu are the differences with relation to the average value of \rmumean for the three energy ranges considered. The differences of \rmumean in all energy intervals is smaller than 1\% for the entire \dx range. The energy independence of \rmumean is expected, since \smutop and \smubur evolve similarly with energy in the range from $10^{17.5}$ to $10^{18.0}$ eV. The lack of energy dependence is an advantage because it allows the analysis of events in a large energy interval, increasing significantly the available statistics. Furthermore, it also contribute to diminishing any effect due to the experimental energy scale.

\begin{figure}
  \centering
  \includegraphics[width=0.7\textwidth]{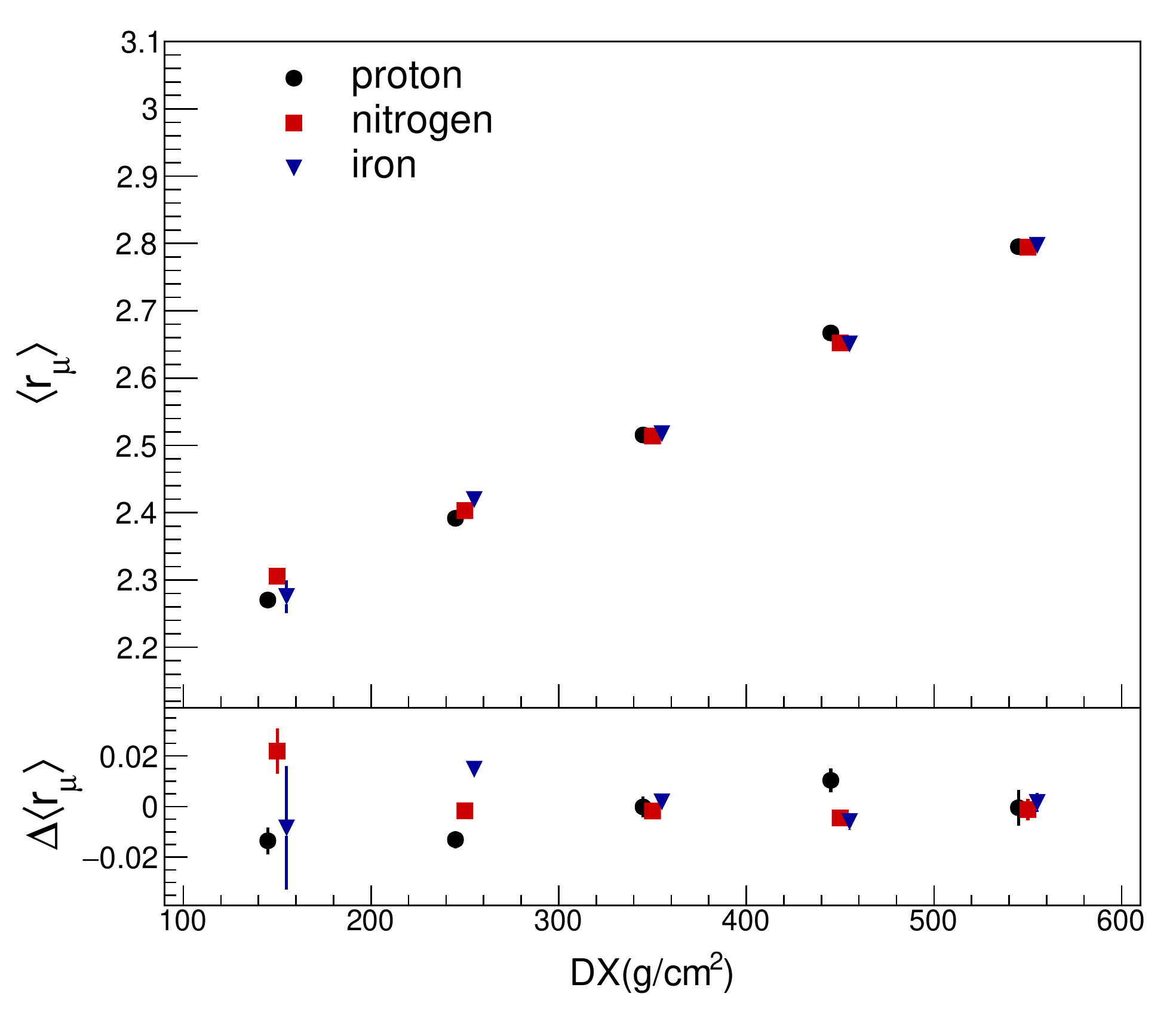}
  \caption{\rmumean as a function of \dx for three primary particles (pr, N, Fe).  All simulated energies are included. Bottom panel shows the difference with relation to the average value. The hadronic interaction model combination used is \qgsjet/\fluka. Points were artificially shifted in \dx for clarity.}
  \label{fig:obser:dx:mass}
\end{figure}

The primary mass dependence of the \rmumean is shown in~\cref{fig:obser:dx:mass}. The hadronic interaction model combination shown is again \qgsjet/\fluka. Very similar results were obtained for all combinations of models. The bottom panel shows the \deltarmumean. The observed primary mass dependence is below  2\%. The dependence of \rmumean with the primary particle was minimized by choosing $\beta=0.6$. The lack of \rmumean dependence on the primary particle is a great advantage because it disantangles the study of the hadronic interaction properties from the determination of the primary particle type.

\section{Results: constraining hadronic interaction models with \rmu parameter}
\label{sec:results}

In this section, the capacity to constrain hadronic interaction models by measuring \rmu is demonstrated by studying the \dx evolution of \rmumean for different combinations of low and high energy hadronic models. Detector resolution is taken into account as explained above. The effect of a limited number of events is considered here. The total number of simulated air showers used (3500) is approximately the number of hybrid events to be measured by the Pierre Auger Observatory infill array in $2$ years of operation. The infill array consists of $750$ m spaced water-Cherenkov stations spread over an area of $23.5$ km$^2$. In this same area the AMIGA-Grande and AugerPrime muon detectors are going to be installed. Considering the duty cycle of the fluorescence detectors to be $14\%$, the total exposure of this experimental set-up is $4.32$ km$^2$.sr.yr. Taking into account the cosmic-rays flux between $10^{17.5}$ and $10^{18.0}$ eV, the expected number of events to be measured per year is 1806.

\begin{figure}
  \centering
  \includegraphics[width=0.7\textwidth]{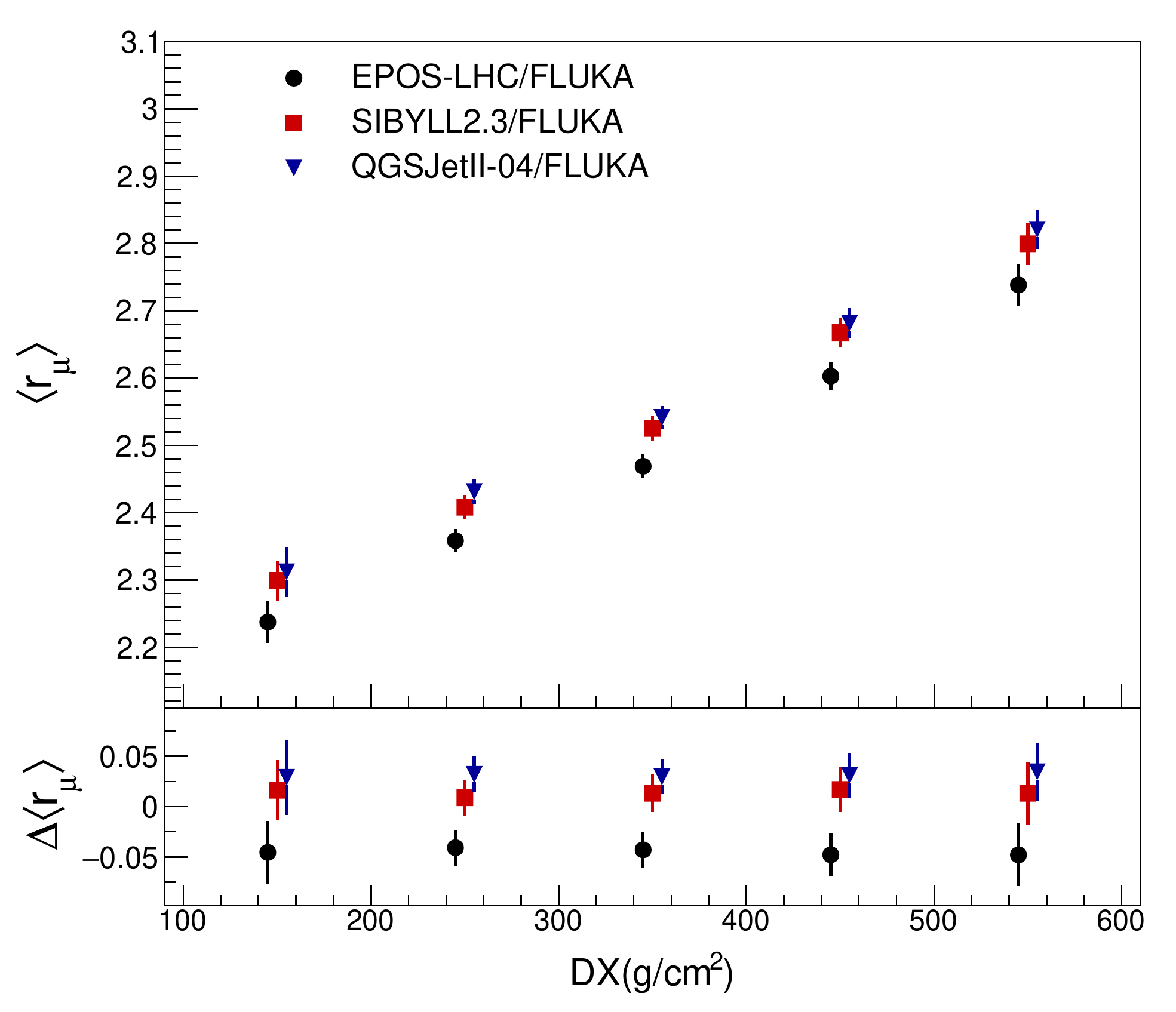}
  \caption{\rmumean as a function of \dx for three high energy hadronic interaction models. All simulated energies are included. All simulated primary particles are included (p, N, Fe). Bottom panel shows the difference with relation to the average value. Low energy hadronic interaction model is \fluka. The detector resolution is set to $20\%$ for \smutop and $10\%$ for \smubur. Points were artificially shifted in \dx for clarity.}
  \label{fig:obser:dx:band:he}
\end{figure}

\begin{figure}
  \centering
  \includegraphics[width=0.7\textwidth]{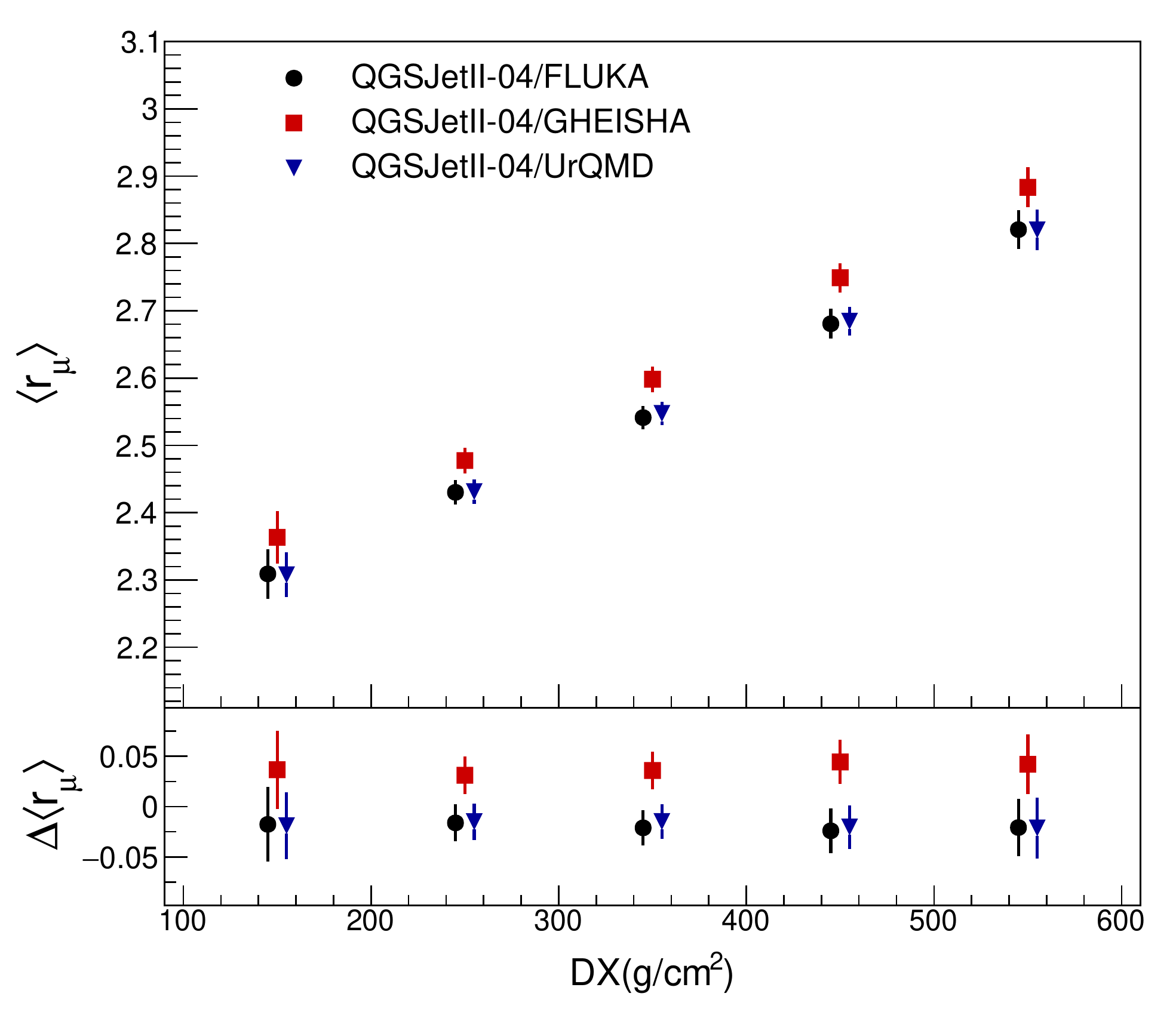}
    \caption{\rmumean as a function of \dx for three low energy hadronic interaction models. All simulated energies are included. All simulated primary particles are included (p, N, Fe). Bottom panel shows the difference with relation to the average value. High energy hadronic interaction model is \qgsjet. The detector resolution is set to $20\%$ for \smutop and $10\%$ for \smubur. Points were artificially shifted in \dx for clarity.}
  \label{fig:obser:dx:band:le}
\end{figure}

\cref{fig:obser:dx:band:he,fig:obser:dx:band:le} show the \rmumean as a function of \dx for different combinations of hadronic interaction models. In~\cref{fig:obser:dx:band:he} the three high energy hadronic models are shown in combination with one low energy hadronic interaction model: \fluka. In~\cref{fig:obser:dx:band:le} the three low energy hadronic models are shown in combination with one high energy hadronic interaction model: \qgsjet. The bottom panels show the \deltarmumean. The worst case for the detector resolution was considered: $10$\% for \smubur and $20$\% for \smutop. \cref{fig:obser:dx:band:he,fig:obser:dx:band:le} show that even with a relatively poor detector resolution a clear separation between hadronic interaction models is achieved. In~\cref{fig:obser:dx:band:he} it is observed that \epos can be distinguished from \qgsjet and \sibyll, while in~\cref{fig:obser:dx:band:le} it is seen that \gheisha can be distinguished from \urqmd and \fluka.

To better quantify the discriminating power of \rmumean, the commonly used Merit Factor can be used. It is defined as:
\begin{equation}
  \textrm{Merit \ Factor} = \frac{\mid \langle r_{\mu}\rangle_a -\langle r_{\mu}\rangle_b \mid}{\sqrt{\sigma_a^2 +\sigma_b^2}},
  \label{eq:merit}
\end{equation}
where $a$ and $b$ refer to any two hadronic interaction model combination and the $\sigma$'s are the standard deviations of \rmumean.

\begin{figure*}
  \centering
    \includegraphics[width=0.47\textwidth]{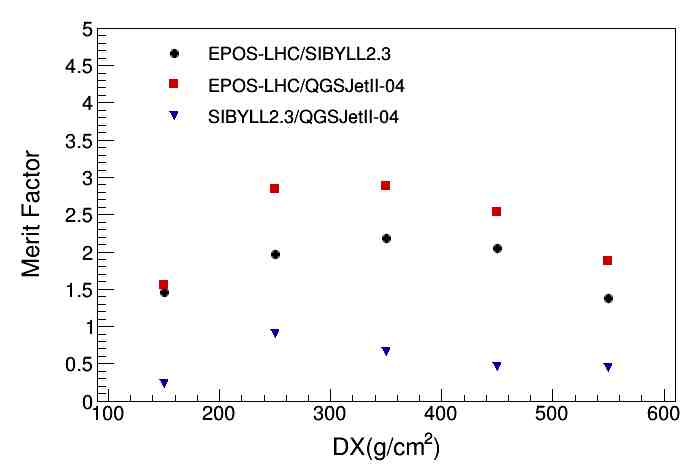}
    \includegraphics[width=0.47\textwidth]{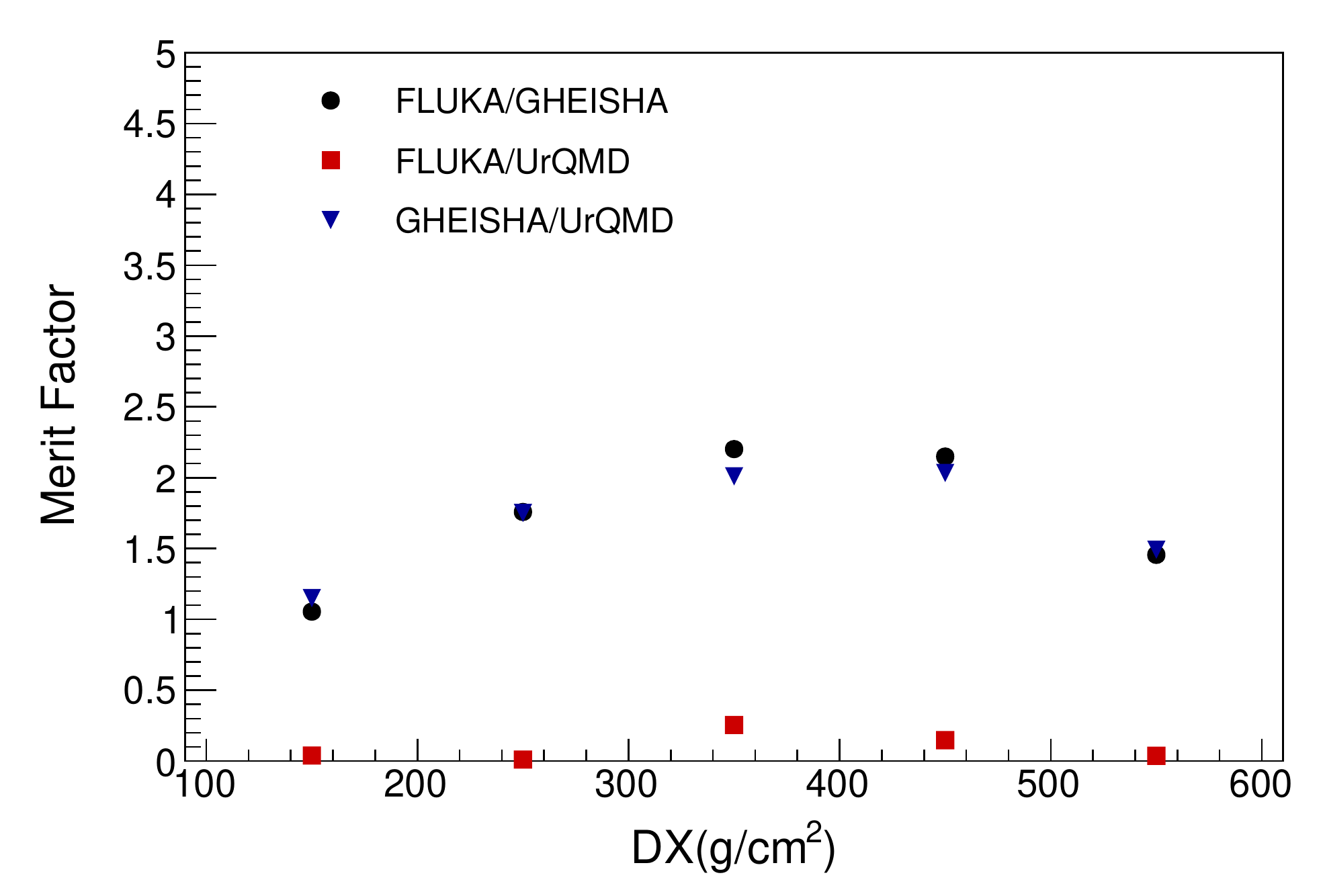}
    \caption{Merit Factor calculated for two different hadronic interaction model combination as a function of \dx. Left panel shows the cases in which the high energy hadronic interaction models are different and the low energy one is the same, \fluka. The legend indicates the two hadronic models considered to calculate the Merit Factor. Right panel shows the cases in which the low energy hadronic interaction models are different and the high energy one is the same, \qgsjet.}
  \label{fig:obser:merit0}
\end{figure*}

\cref{fig:obser:merit0} shows the Merit Factor as a function of \dx for the same experimental resolutions and statistics described above. The left-hand panel refers to hadronic model combinations with different high energy models, and on the right-hand panel with different low energy models. The best Merit Factor is $3.0$ between \epos and \qgsjet and $2.3$ between \fluka and \gheisha.

\begin{figure*}
  \centering
  \includegraphics[width=0.32\textwidth]{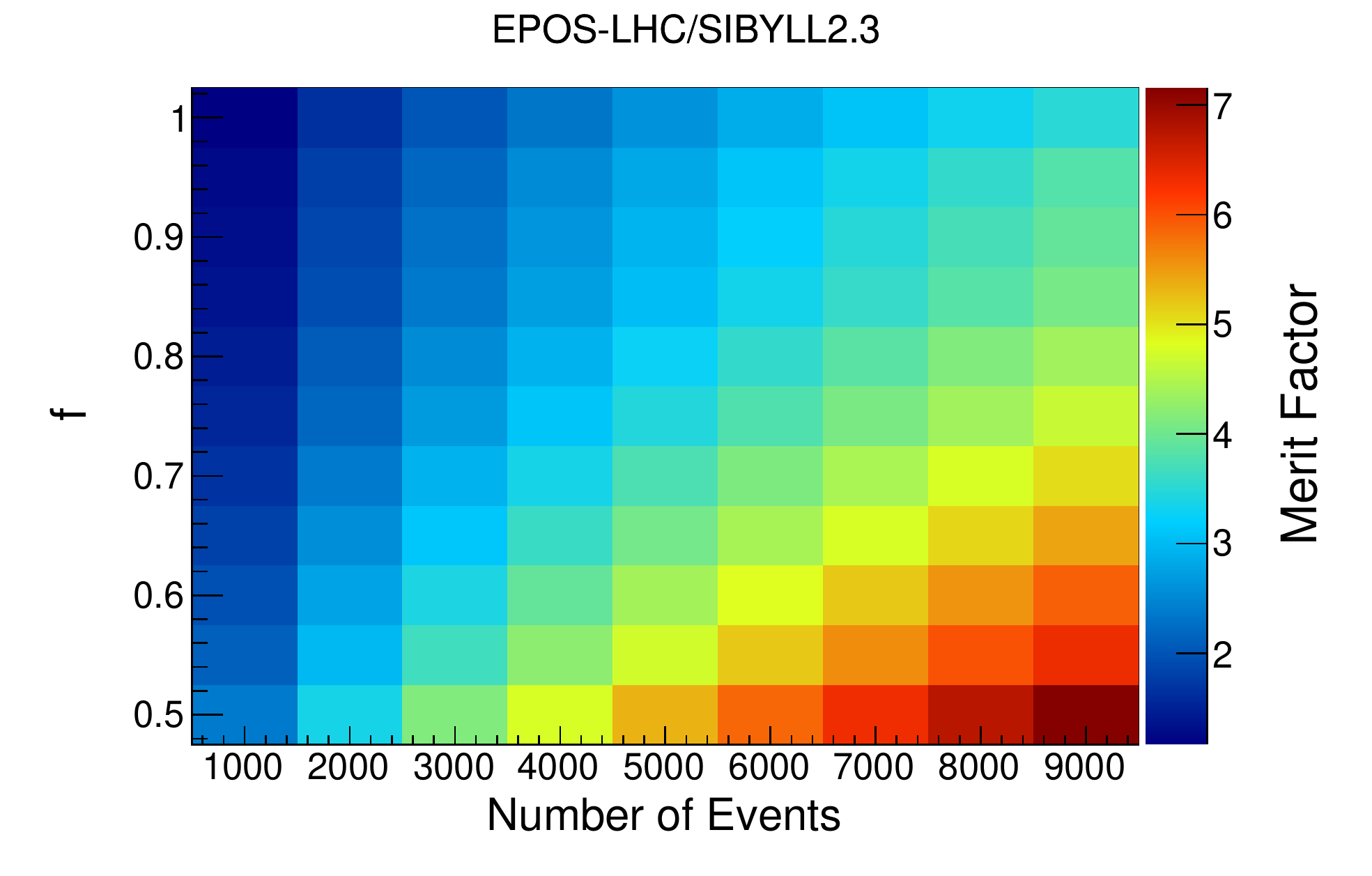}
  \includegraphics[width=0.32\textwidth]{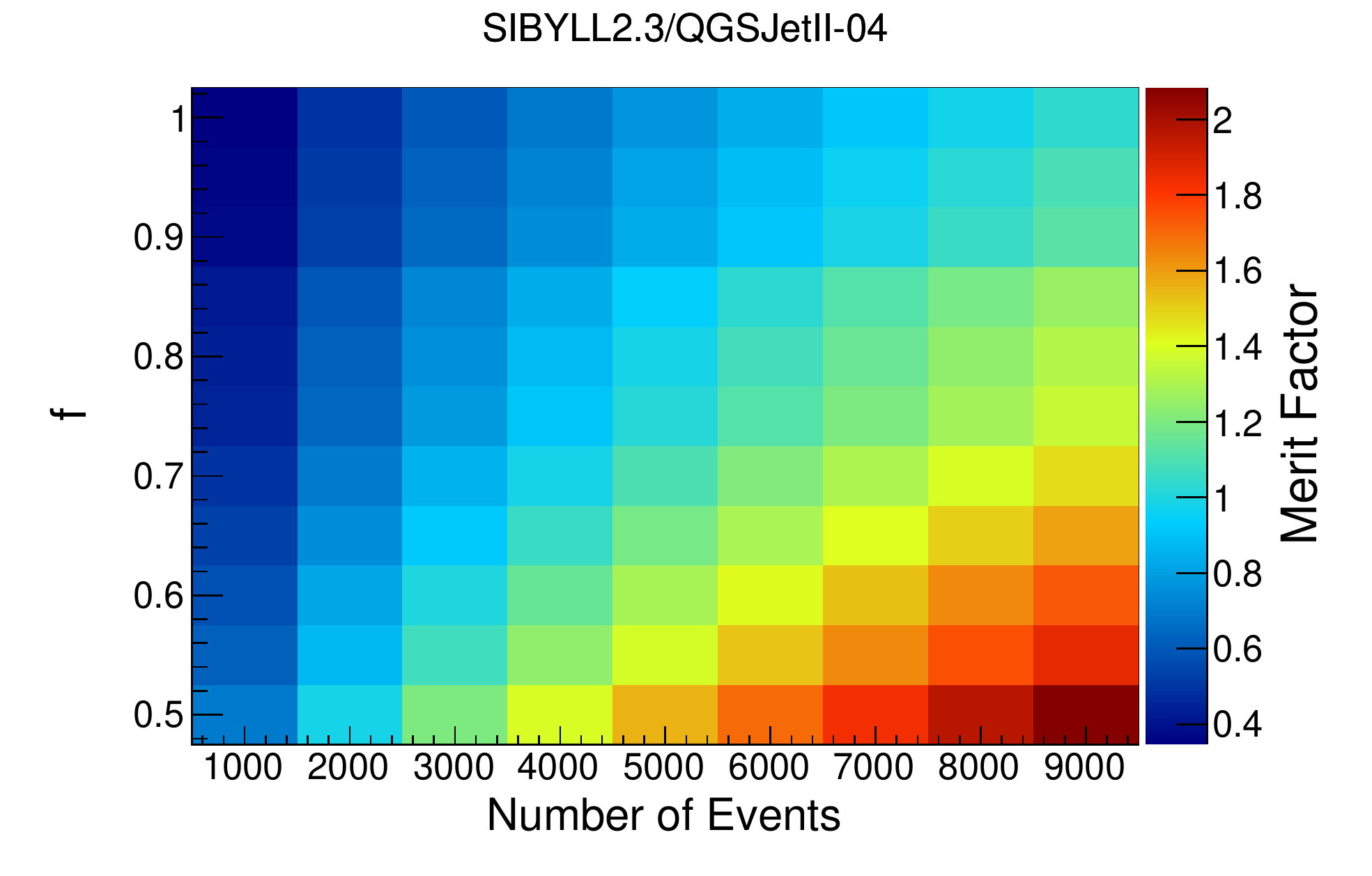}
  \includegraphics[width=0.32\textwidth]{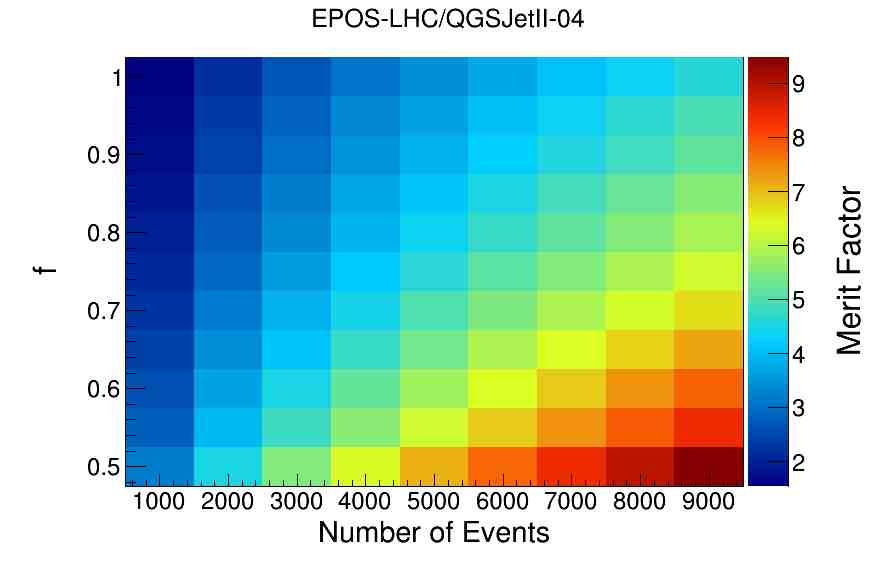}
  \caption{Merit Factor as a function of the number of events and detector resolution. Detector resolution are defined as $\sigma_{bur} = 0.1 f$\smubur and $\sigma_{top} = 0.2 f$\smutop. The Merit Factor is given in the color scale.}
  \label{fig:obser:merit:he:3d}
  \end{figure*}

\begin{figure*}
  \centering
  \includegraphics[width=0.32\textwidth]{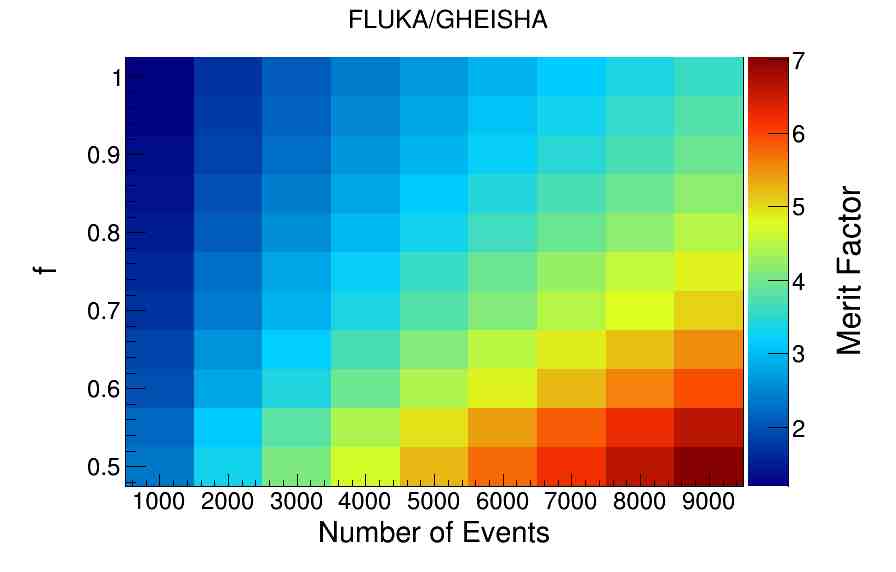}
  \includegraphics[width=0.32\textwidth]{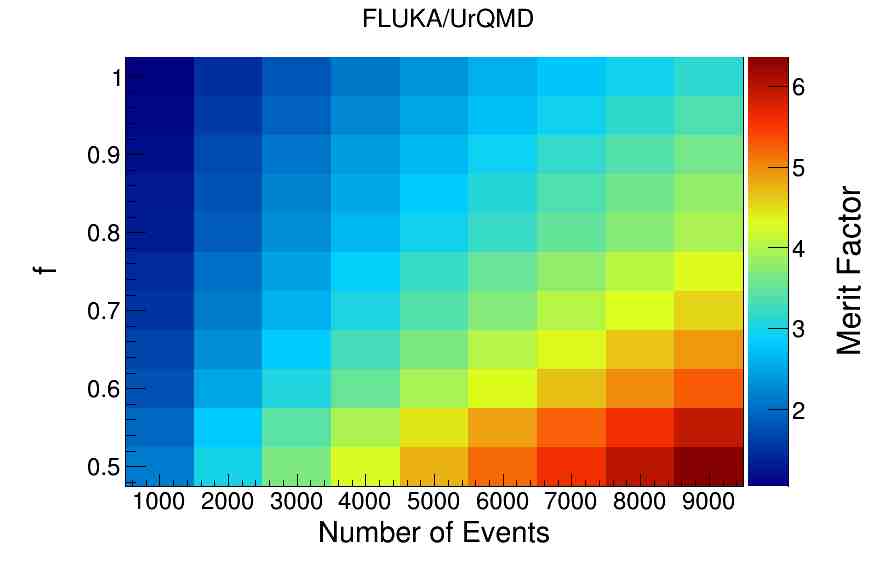}
  \includegraphics[width=0.32\textwidth]{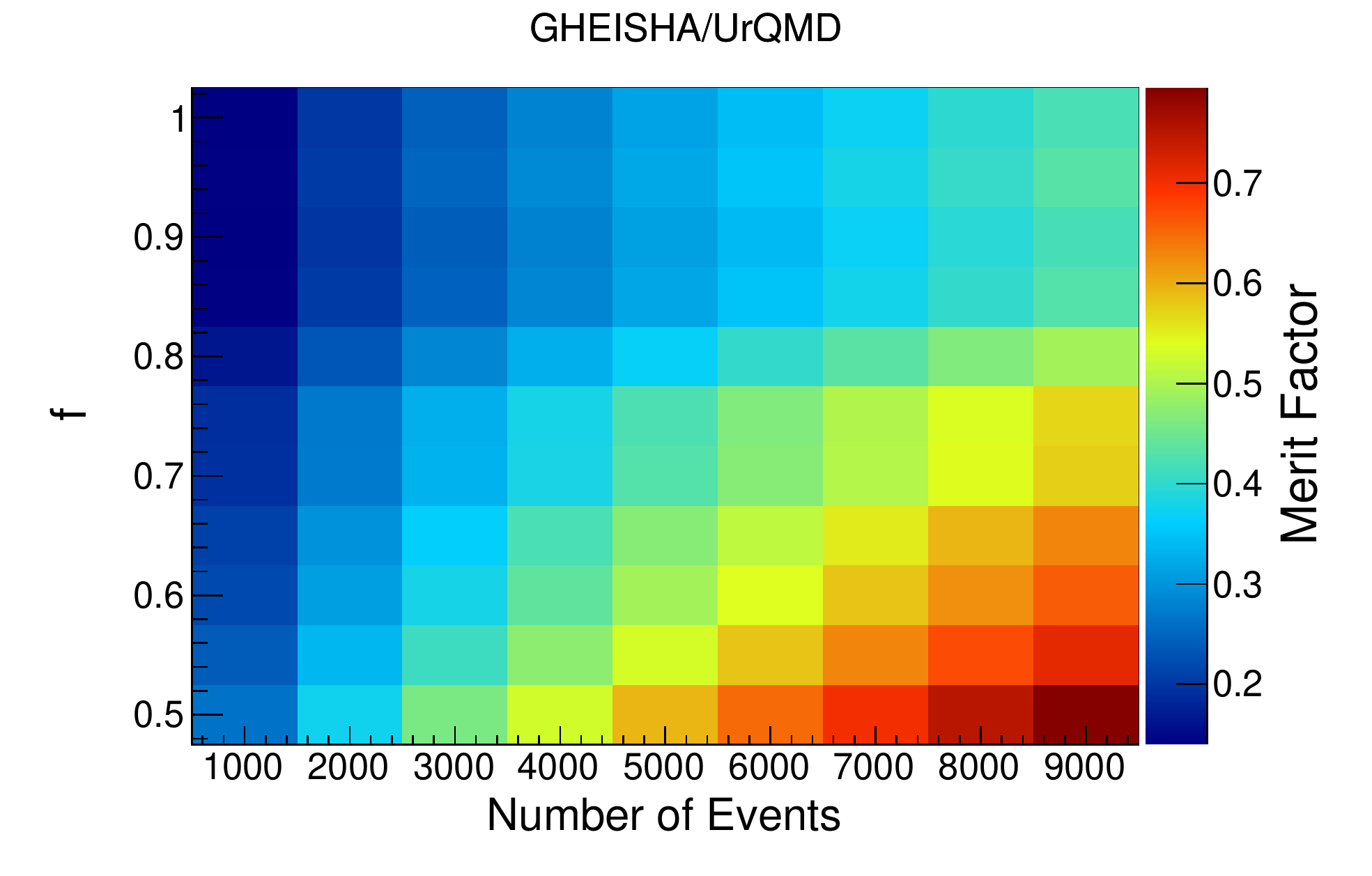}
  \caption{Merit Factor as a function of the number of events and detector resolution. Detector resolution are defined as $\sigma_{bur} = 0.1 f$\smubur and $\sigma_{top} = 0.2 f$\smutop. The Merit Factor is given in the color scale.}
  \label{fig:obser:merit:le:3d}
  \end{figure*}

\cref{fig:obser:merit:he:3d,fig:obser:merit:le:3d} show the Merit Factor as a function of the number of events and detector resolution for one particular \dx bin: $300 < DX/(\rm{g/cm}^2) < 400$. The resolutions on \smutop and \smubur were re-scaled by a common factor $f$, being that $\sigma_{\rm{bur}} = 0.1 f S_{\mu}^{\rm{bur}}$ and $\sigma_{\rm{top}} = 0.2 f S_{\mu}^{\rm{top}}$. The effect of the number of events was calculated by re-scaling the standard deviation of the muon densities by a factor $\sqrt{N_{\rm{sim}}/N}$, where $N_{\rm{sim}}$ is the number of simulated showers and $N$ is the number of showers in each case.

\cref{fig:obser:merit:he:3d} shows that it is possible to reach large Merit Factor values ($> 5$) for the separation between \epos and \qgsjet/\sibyll using a reasonably small number of events ($< 6000$) considering realistic detector resolutions ($\sigma_{\rm{bur}}/S_{\mu}^{\rm{bur}} < 0.06$ and $\sigma_{\rm{top}}/S_{\mu}^{\rm{top}} < 0.13$). On the other hand, the separation between \sibyll and \qgsjet is small for any resolutions and number of events, which is expected because of their similar values of $\eta$. The same conclusions can be drawn about the low energy hadronic interactions models from \cref{fig:obser:merit:le:3d}. \rmumean provides a very good separation between \fluka and \gheisha/\urqmd, but the separation power is limited for \gheisha and \urqmd.

\section{Conclusions}
\label{sec:conclusion}

This paper studies the ground muon energy spectrum of air showers and proposes an analysis procedure to constrain hadronic interaction models used in Monte Carlo simulations. In~\cref{sec:carac}, it was shown that the energy distribution of muons at ground level can be well described by an asymmetric Gaussian function with four parameters. The study of the evolution of the four parameters with \dx concluded that the overall shape of the energy spectrum of muons does not depend strongly on the combination of low and high energy hadronic interactions models. Also, it was verified that the average muon energy, or the parameter $\eta$, is sensitive to the model combination and presents a strong evolution with \dx.

However, $\eta$ is not an easy parameter to be measured. Therefore in~\cref{sec:obser} a new and experimentally motivated parameter (\rmu) is proposed and its correlation with $\eta$ is shown. \rmu dependencies on the primary mass and energy were proven to be insignificant which allows on to disentangle the composition and the hadronic interaction studies as well as minimize the effect of systematic uncertainties in the energy reconstruction.

The general properties of the current and proposed muon detectors of the Pierre Auger Observatory are considered to study \rmu under realistic experimental limitations. \cref{sec:results} shows that the discrimination power of \rmu is significantly large. \epos can be separated from \sibyll and \qgsjet with large Merit Factor ($> 5$) using a reasonably small number of events ($< 6000$). \sibyll and \qgsjet show similar average muon energy at ground and therefore can, in the best case, be discriminated with Merit Factor $\sim 2$ with about $9000$ events. As for the low energy hadronic models, \fluka can be separated from \gheisha and \urqmd with large Merit Factor ($> 5$) using a reasonably small number of events ($< 6000$). \gheisha and \urqmd can, in the best case, be discriminated with Merit Factor $\sim 0.8$ with about $9000$ events. It was shown that correlated systematic uncertainties on \smutop and \smubur have insignificant influence on \rmumean. Uncorrelated oppositive systematics of order of 2.5\% can generate a systematic uncertainty on \rmumean of the same order of the separation betweem the hadronic interaction models. This means that for a realistic application of \rmu analysis, the systematic uncertainties on the muon densities measured by the two detectors have to be correlated (more realistic case). If the systematic uncertainties of the two detectors are in oppositive directions (less realistic case) they have to be smaller than 2.5\% in order to keep the discrimination power of the hadronic interaction models.

It was also shown that \rmu is a very robust parameter which can be used irrespective of ignorance of the primary mass composition to test the hadronic interaction models. Constrains imposed by an analysis based on \rmu can in a short period of time contribute to the solution of the know problems with muon production in extensive air shower~\cite{Aab2015a,MuonsPRL}.

\section*{Acknowledgements}

The authors would like to thank Roger Clay for the review of the manuscript on behalf of the Pierre Auger Collaboration. RRP thanks the financial support given by FAPESP (2014/10460-1).  VdS thanks the support of the Brazilian population via CNPq and FAPESP (2015/15897-1, 2014/19946-4).

\section*{References}
\bibliographystyle{elsarticle-num}
\bibliography{mylib.bib}

\newpage


\end{document}